# Solar System Abundances and Condensation Temperatures of the Halogens Fluorine, Chlorine, Bromine, and Iodine

Katharina Lodders and Bruce Fegley, Jr.

Planetary Chemistry Laboratory, Department of Earth & Planetary Sciences and McDonnell Center for the Space Sciences, Washington University, One Brookings Drive, Saint Louis MO 63130 U.S.A.

**Abstract:** We review a large body of available meteoritic and stellar halogen data in the literature used for solar system abundances (i.e., representative abundances of the solar system at the time of its formation) and associated analytical problems. Claims of lower solar system chlorine, bromine and iodine abundances from recent analyses of CI-chondrites are untenable because of incompatibility of such low values with nuclear abundance systematics and independent measurements of halogens in the sun and other stars. We suspect analytical problems associated with these peculiar rock types have led to lower analytical results in several studies. We review available analytical procedures and concentrations of halogens in chondrites. Our recommended values are close to previously accepted values. Average concentrations by mass for CI-chondrites are F = 92±20 ppm, Cl = 717±110 ppm, Br = 3.77±0.90 ppm, and I = 0.77±0.31 ppm. The meteoritic abundances on the atomic scale normalized to N(Si) =$10^6$ are N(F) = 1270±270, N(Cl) = 5290±810, N(Br)= 12.3±2.9, and N(I) = 1.59±0.64. The meteoritic logarithmic abundances scaled to present-day photospheric abundances with log N(H) = 12 are A(F)= 4.61±0.09, A(Cl)=5.23±0.06, A(Br) = 2.60±0.09, and A(I) = 1.71±0.15. These are our recommended present-day solar system abundances. These are compared to the present-day solar values derived from sunspots of N(F) = 776±260, A(F) = 4.40±0.25, and N(Cl) = 5500±810, A(Cl) = 5.25±0.12. The recommended solar system abundances based on meteorites are consistent with F and Cl abundance ratios measured independently in other stars and other astronomical environments. The recently determined chlorine abundance of 776±21 ppm by Yokoyama et al. (2022) for the CI-chondrite-like asteroid Ryugu is consistent with the chlorine abundance evaluated for CI-chondrites here. Historically, the halogen abundances have been quite uncertain and unfortunately remain so. We still need reliable measurements from large, representative, and well-homogenized CI-chondrite samples. The analysis of F, Br, and I in Ryugu samples should also help to obtain more reliable halogen abundances. Updated equilibrium 50% condensation temperatures from our previous work (Lodders 2003, Fegley & Schaefer 2010, Fegley & Lodders 2018) are 713 K(F), 427 K (Cl), 392 K (Br) and 312 K (I) at a total pressure of $10^{-4}$ bar for a solar composition gas. We give condensation temperatures considering solid-solution as well as kinetic inhibition effects. Condensation temperatures computed with lower halogen abundances do not represent the correct condensation temperatures from a solar composition gas.

**Keywords**: abundances, sun, meteorites, solar system, condensation temperatures, fluorine, chlorine, bromine, iodine

## 1. Introduction

This work is an update to Lodders (2003) for the halogen abundances and their 50% condensation temperatures since new measurements on halogen abundances and revised thermodynamic properties have become available. We adopt the same procedures as in Lodders (2003) for our evaluation and calculations. These updates also apply to Lodders et al. (2009), and Palme et al. (2014). The halogens fluorine, chlorine, bromine, and iodine belong to the volatile elements, and are difficult to measure in the solar photosphere. The abundances of F and Cl can be derived from sunspot spectra, but Br and I solar system abundances are derived only from meteorite measurements.

The elemental concentrations of the five observed falls of the carbonaceous CI-chondrites Alais (6 kg preserved, France 1806), Ivuna (0.7 kg, Tanzania 1938), Orgueil (14 kg, France 1864), Revelstoke (<1 g, Canada 1964), Tonk (10 g. India 1911) are frequently used as a solar system abundance standard for elements that do not form gases under typical terrestrial conditions. The CI-chondrites have the highest degree of aqueous alteration and therefore are of petrological type 1 (CI1). (The Tagish Lake chondrite fall was originally considered as a CI-chondrite of petrological type 2 (CI2) but is now classified as a C2-ungrouped chondrite because its volatile elements are fractionated when compared to the CI-chondrite group).

Data for halogens in meteorites, and in particular, CI-chondrites are sparse, and sometimes abundances in other carbonaceous chondrites (such as CM, CV chondrites) or interpolations from elemental abundance distribution curves have been used to constrain halogen solar system abundances see, e.g., Anders & Ebihara, (1982), Anders & Grevesse (1989). In section 2 we describe the challenging halogen measurements in rocks and meteorites. In section 3, we discuss the available meteorite data and our recommended solar system halogen abundances. Section 4 compares the meteoritic data to measurements of halogens in the sun and other astronomical environs. Section 5 discusses our updates to condensation temperatures and our currently recommended values. We give conclusions in section 6.

## 2. Halogens in Chondrites

Critical assessment and evaluation of halogen analyses must consider two interrelated issues: The analytical methods used and the nature of the halogen-bearing phases in the meteorites. Halogen analyses often require quantitative chemical separation of halogens from host rocks. As described in the following sections water soluble, insoluble, and organic halogen-bearing phases are present in meteorites The chemical separation method depends on the nature of the host phases for the halogens and therefore, sample treatment and preparation can affect quantitative results.

The comprehensive review by Brearley & Jones (2018) provides a good summary about halogens in meteorites, and we focus on some basic and additional information here. The CI-chondrites are rare and we consider the halogens in other carbonaceous chondrites to obtain a better understanding of the halogen-bearing phases. Meteorites contain halogens in silicates, phosphates, salts, and organic compounds. A more unusual host phase is troilite where Mason and colleagues (Allen & Mason 1973, Mason & Graham 1970) detected bromine and iodine.

Salts and organics are important halogen carriers in the CI- and other carbonaceous chondrites. While halite (NaCl) and sylvite (KCl) crystals occur in the Zag and Monahans ordinary chondrites (Gibson et al. 1998, Zolensky et al. 2000, Rubin et al. 2002; Bridges et al. 1997, 2004) and Barber (1981) found Br-bearing halite (NaCl) in the Murchison CM2 chondrite, fine-grained halogen salts may be hard to detect in CI-chondrites. We speculate that organic coatings or "films" on existing grain surfaces could impede salt grain detections, e.g., see the electron microscopy studies of carbonaceous coatings on olivine and FeS grains in the Allende CV3 chondrite (Bauman et al. 1973, Green et al 1971, Harris and Vis 2003). Some organics carry halogens such as the chlorinated aromatic hydrocarbons reported by Studier et al. (1965) and Hayes & Biemann (1968). Müller (1953) recorded 4.8 percent chlorine in the organic matter extracted by solvents from Cold Bokkeveld (CM2), and Schöler et al. (2005) found F and Br in such extracts in this chondrite. In this context the discovery of organo-halogens (i.e., $CH_3Cl$) in comets and protostellar disks is also noteworthy (Fayolle et al. 2017) making it plausible that the organo-halogens in meteorites are indigenous and not a contamination product. Both CI and CM chondrites experienced aqueous alteration on their parent bodies. Water soluble phases are mainly sulfate and alkali halide salts, while the carbonates are usually much less soluble. Lawrencite, $FeCl_2$, has been putatively identified in the Indarch EH chondrite (Keil 1968) but the experiments by Muenow et al (1992) rule out lawrencite as a major carrier of Cl. Instead, halogens in EH chondrites are often concentrated in mineral phases such as djerfisherite, $K_3Cu(Fe,Ni)_{12}(S,Cl)_{14}$ (Fuchs 1966, Keil 1968) which contains up to 1.5% Cl in EH chondrites (El Goresy et al. 1988).

Among the carbonaceous chondrites, indigenous alkali and ammonium halide and sulfate salts were extractable with water in freshly collected samples of the Orgueil CI-chondrite (Cloëz 1864, Pisani 1864). Ammonium salts have not been reported in recent studies presumably because they were either not looked for or because ammonia was lost from the rock while stored in museums. (However, ammonia is released when kerogen-like insoluble matter isolated from carbonaceous chondrites is heated, Pizzarello & William 2012). Leaching experiments on carbonaceous chondrites typically indicate dissolution of NaCl and KCl but these salts have not been directly observed yet in CI-chondrites. Several studies report fractions of water-soluble and non-soluble (= structurally bound) halogens in carbonaceous chondrites (e.g., Goles & Anders 1962, Reed & Allen 1966, Tarter et al. 1980, Moore et al. 1982, Bonifacie et al. 2007, Sharp et al. 2007, 2013). Leaching experiments on pristine (least terrestrially altered, if any) samples of the Tagish Lake meteorite (fall 1999) confirmed earlier results that Mg, Ca, Na, and K sulfates and chlorides are readily leached from carbonaceous chondrites (e.g., Izawa et al. 2010). The recently fallen Aguas Zarcas CM2 chondrite contains 0.6% water-soluble salts; the recrystallized salts were identified as: halite (NaCl), chlorartinite ($Mg_2(CO_3)(OH)Cl\cdot 2H_2O$), thenardite ($Na_2$SO4), and sodium chlorate ($NaClO_4$) (Garvie 2021).

The presence of water-soluble halides in meteorites raises the issue whether terrestrial contamination with or loss of salts has occurred. This is a possibility in particular for meteorite finds and contamination with selected elements is known for chondrites collected from deserts (e.g., contamination with Ca-sulfates) and in Antarctica (iodine but always in Br or Cl; Dreibus et al. 1986, Heumann et al. 1987). However, the reverse, loss of soluble halogen salts during weathering seems to be a larger issue when analyzing meteorite finds. However, gain or loss of halogens and other elements in meteorites collected shortly after they fell is far less probable, although contamination or less can never be fully ruled out. For example, Daubrée (1872) reported 0.12% water soluble NaCl in the Lance CO3 carbonaceous chondrite (fall 1872). This corresponds to 730 ppm chlorine in the whole-rock. His sample had been in the ground for 3 days but he did not find other soluble salts (such as Ca-bearing salts) indicative of terrestrial contamination. His sample was covered completely by fusion crust which sealed the meteorite from soil which was reported as dry, and there was no rain between the time of the meteorite fall and its collection. Under $H_2$ –flow and red heat the stone released a sublimate of NaCl in the same proportions as found in the water extract. In a footnote he states" Ce n'est d'ailleurs pas la premiere fois que du chlorure de sodium est signale dans des meteorites".

Loss of halogen salts must be considered when meteorite samples are rinsed to remove potential "surface contamination" before analysis, which was done in several studies (e.g., Magenheim et al. 1994, Bonifacie et al. 2007, Sharp et al. 2013). However, while halogen losses by rinsing with even cold water could be important for many meteorite samples, rinsing of CI-chondrites is not possible because they immediately fall apart in contact with water and cannot be handled like other solid samples (e.g., Berzelius 1834).

The following is particularly important for the CI-chondrites because they contain up to 20 percent of indigenous water in hydrated silicates and as crystal-water in salts (e.g., epsomite $MgSO_4 \cdot 7H_2O$). Epsomite and hydrated silicates release their bound water upon sample heating to a few hundred degrees, and will re-gain water upon cooling in air. The mobilization of water may leach salts and water-soluble organics, and cause re-distribution of halogen bearing compounds within the rock. Heating also can release non-water soluble halogen-bearing organics at elevated temperatures, and salts are certainly lost when samples are fused.

Phosphates, in particular Cl-rich-apatite (5-6% Cl) with minor F substitution [$Ca_5(PO_4)_3(OH,F,Cl)$], are essentially secondary metamorphic halogen-bearing phases in chondrites (Fuchs 1969, Lewis & Jones 2016, Ward et al. 2017, Brearely & Jones 2018). Fluorapatite, an expected thermodynamically stable condensate host phase for fluorine (see below), is extremely rare in meteorites but has been reported (with up to 5.23 mass% F) in exotic graphite-rich xenoliths found in the ordinary chondrite Krymka which seem to have a carbonaceous-chondrite-like precursor (Semenenko et al. 1995, 2004).

The CI-chondrites also contain phosphates. Jungck et al. (1981) and Meier et al. (1981) suspected the presence of hydroxy- or fluorapatite in the Orgueil CI-chondrite but ruled out chlorapatite. Element-mapping in CI-chondrites revealed the association of fluorine with Ca-phosphate, particularly at the rims, but Morlok et al. (2006) did not report the F concentrations and presence of Cl in these phases.

Other halogen-rich phases have been found. Boström & Fredriksson (1966) reported up to 3 mass% chlorine in the alteration phase limonite [$FeO(OH) \cdot nH_2O$] in the CI-chondrite Orgueil. Although not found in CI-chondrites, unequilibrated ordinary chondrites (type LL3) contain chlorine in the feldspathoid sodalite with 6.4-7.18% (by mass), 2.0 – 5% in scapolite, 0.9-2.1% in nepheline, 0.7-4.6% in glassy mesostasis, and possibly up to 0.2% in plagioclase (Alexander et al. 1987, Bridges et al. 1997). Wasserburg et al. (2011) found 7.1% Cl in a Na-Cl rich chondrule from Allende (CV3). Grossman et al. (1979) reported percent-level Cl-contents in an Amboeboid Olivine Aggregate (AOA) from Allende. Sodalite, an expected Cl-bearing condensate at low temperatures, has been found in calcium-aluminum-rich inclusions (CAIs) from carbonaceous chondritesIn particular, the fine-grained Inclusions (FGI) are important halogen and alkali element carriers, e.g, Spettel et al. (1978) found that concentrations of Na, Cl and Br appear to be correlated. Quijano-Rico & Wanke (1969) found what seems to be a FGI {their chondrule C) with 7.3 % Cl, (see discussion in Clarke et al. 1970). Zaikowski (1980) found large enrichments in iodine in FGI. The origin of halogens in CAIs is debated but already Wänke et al. (1974) and Grossman & Ganapathy (1975) make plausible arguments that sodalite is likely a secondary alteration product (see more discussion in condensation section below).

In addition to analytical uncertainties and natural variations of halogens within small analyzed samples noted in section 2.3, there are apparent variations of halogen concentrations between the known CI-chondrites Alais, Ivuna and Orgueil. For example, Ivuna has higher Br contents than Orgueil and Alais (e.g., Anders & Ebihara 1982, Burnett et al. 1989), and Reed & Allen (1966) noticed that Orgueil contains more leachable Cl than Ivuna. Orgueil fell in 1864 and Ivuna in 1935, and it is known that terrestrial humidity affects the salt chemistry. It cannot be ruled out that cation exchange of sulfates with halides affected aqueous solubilities and halogen redistribution as a function of terrestrial residence time. Formation of white sulfate efflorescence was seen within short times after the CI-chondrites fell (Cloëz 1864, Daubrée 1864a,b; for Orgueil; Berzelius 1834 for Alais) and sulfate formation from sulfides is ongoing in these meteorites (for a quantitative description see Lodders & Fegley 2011).

It is unlikely that analytical results for high halogen concentrations can all be ascribed to sample contamination if different samples of a given meteorite from different sources were analyzed. Contamination in one group's lab cannot explain why different groups would obtain similar results, which instead could point to contamination of samples at the source. On the other hand, there is some indication that studies giving systematically lower values involve more elaborate chemical analysis, as described below.

## 2.1. Analytical Methods

Most available measurements are done by neutron activation analyses (NAA) combined with chemical separation of halogens after irradiation. Other quantitative measurements are done with mass-spectrometry, ion chromatography, and for fluorine, with ion-selective electrode. In this section we describe the chemical separation techniques (2.2.1) and NAA (2.2.2) and discuss other analytical techniques specific to a given halogen in the separate element sections below.

Sample size is important if minute minerals are present that sequester halogens. Small randomly selected samples of CI-chondrites are known to have a wide compositional heterogeneity as illustrated by Greshake et al. (1998) and Morlok et al. (2006). Greshake et al. (1998) analyzed 50-micrometer size particles from Alais and Orgueil by proton-induced X-ray emission (PIXE) spectroscopy and Figure 1 shows the frequency distribution for their chlorine concentrations in 200-ppm intervals. Since the samples were very small, the analyses are normalized to the mean CI-chondrite concentration of 18.66 wt% iron in Palme et al. (2014). The spread in chlorine values is considerable but not unexpected because CI-chondrites are not homogenous on the 50-micrometer scale.

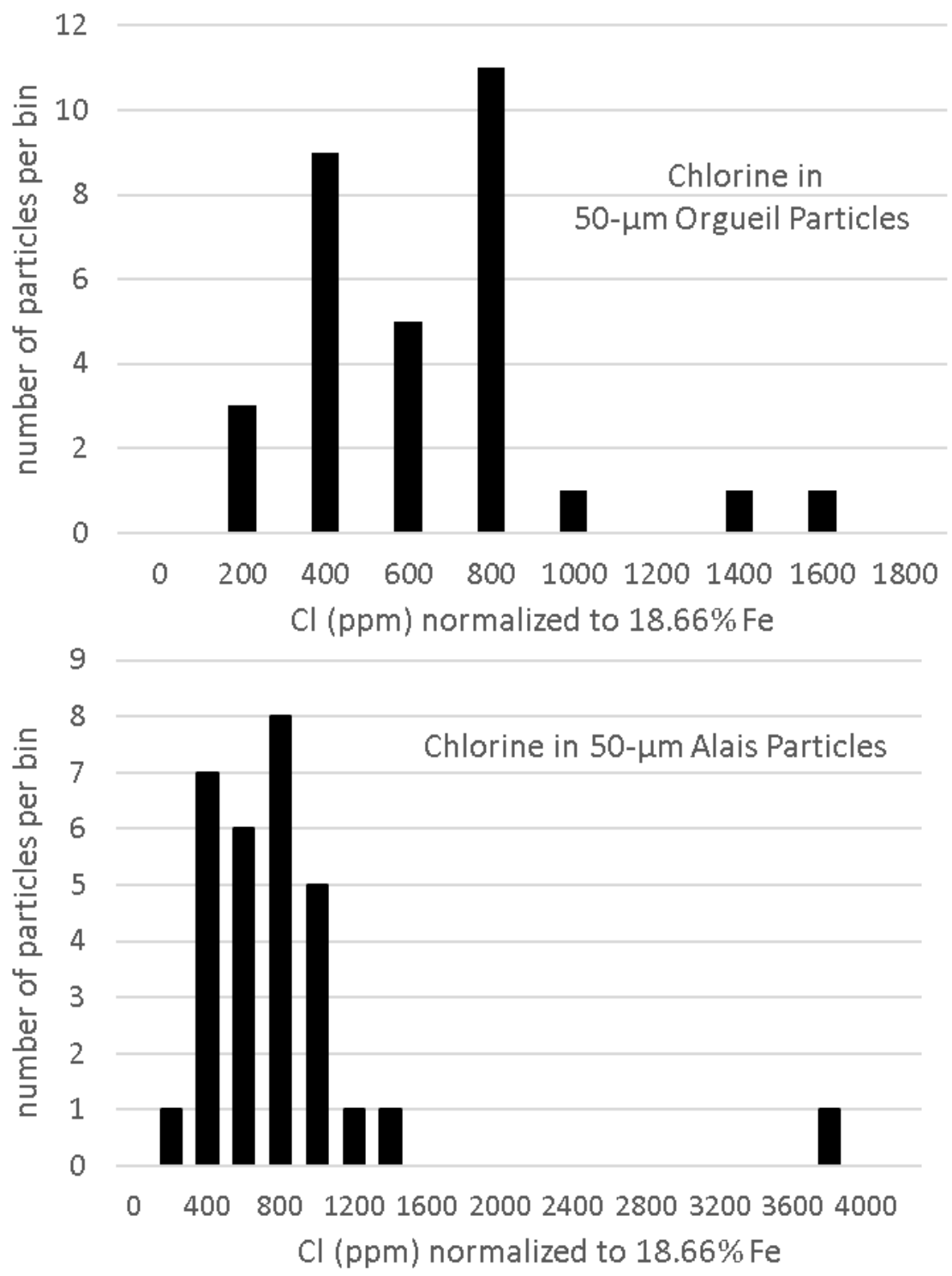

**Figure 1**. Frequency of chlorine concentration of 50-micrometer size samples from Alais and Orgueil. The concentration bin step size is 200 ppm. Data from Greshake et al. (1998).

Morlok et al. (2006) estimated that the mass "threshold" for chemical heterogeneities is smaller than 1-2 g for CI-chondrites. Therefore, bulk analyses are ideally done on aliquots from several grams of homogenized meteorite samples. Analyses of larger 50-200 mg aliquots or samples were used by several authors for elemental bulk analyses, but samples from CI-chondrites are limited due to the rare occurrence of these meteorites. Larger sample sizes of 100-200 mg should be preferred for representative bulk analyses, and extremely small samples sizes of 1-2 milligrams, as used in the study by Clay et al. (2017) may explain some of their very low halogen concentrations found in comparison to other studies as described below.

The draw-back of sample homogenization is that contaminants are easily introduced. Contamination is also an issue when reagents are needed during chemical processing for analyses, which can be quite extensive for halogen analyses. However, reagent contamination is not an issue for samples that are analyzed by neutron activation because chemical processing occurs after the samples have been irradiated because only activated nuclei are measured (see below).

#### 2.1.1. Chemical Separation/Extraction Procedures

Many analysis techniques require some chemical separation of the halogens to increase sensitivity because their abundances in meteorites are relatively low. Although results of non-destructive instrumental neutron activation analyses should not be dependent on the host phase of the halogens, the chemical and physical properties of the halogen host phases can affect results if the halogens (and/or their decay products) can escape from the samples and are not completely recovered from the sample container used during irradiation (see 2.1.2 below). Chemical processing always has the potential for halogen loss and contamination from reagents. Volatile and water-soluble halogen compounds in the samples may easily be lost during sample handling and cleaning, open sample fusion, or leaching. On the other hand, most separation methods exploit the solubility and volatilization of the halogens and require quantitative re-capture of volatilized compounds (see, e.g., Anders 1964; Krähenbühl et al. 1973a,b; Goles et al. 1967; Nakamura et al. 2011).

Typical separation methods include sample fusion with $Na_2O_2$-NaOH (e.g., Reed & Allen 1966; Goles et al. 1967; Liebermann & Ehmann 1967; Krähenbühl et al. 1973a,b; Ebihara et al. 1982) and extracting halogens with steam from melted samples (pyrohydrolysis; e.g., Dreibus et al. 1979; Sharp et al. 2007, 2013; Menard et al. 2013), ion chromatography (e.g., Magenheim et al. 1994; Bonifacie et al. 2007; Nakamura et al. 2011; Menard et al. 2013; Sharp et al. 2013) or laser-heating for noble gas release after neutron irradiation (e.g., Clay et al. 2017).

Sample fusion with NaOH or $Na_2O_2$-NaOH flux and leaching the fusion cake with water brings the halogens into solution for further processing. To isolate the halogens from other salts such as sulfates, subsequent distillation under $N_2$ flow selectively forces the halogens out of solution, and they are then trapped in water (e.g., Greenland 1963; Greenland & Lovering 1965; Von Gunten et al. 1965; Wyttenbach et al. 1965; Ozaki & Ebihara 2007). After irradiation in neutron activation analyses carrier salts are often added to the sample and the flux before fusion to enable more quantitative halogen capture. The addition of carriers also allows the determination of the yields of the chemical procedures.

"Dry" fusion (> 1200 C) of samples with $Na_2O_2$-NaOH causes volatilization of alkali halides (e.g., NaCl, NaBr). This method is similar to that used by Edwards & Urey (1955) to extract alkali elements by heating samples with addition of $CaCl_2$. In order to ensure quantitative extraction of alkalis, they added excess chloride; and for quantitative extraction of halogens, addition of excess sodium compounds seems to be advised.

Pyrohydrolysis is a commonly applied extraction method for halogen analyses. The setup typically consists of a gas flow apparatus made of silica where gas passes over a sample that is heated with an induction coil. Sample fusion with $Na_2O_2$-NaOH under a flow of steam facilitates the removal of halogen-bearing gases which are usually trapped in NaOH solution. The actual gas species driven out of the fused samples are not really known but thermodynamic calculations indicate that pyrohydrolysis likely leads to volatilization of alkali halides (NaX, KX) and other volatilized compounds such as HX, and possibly $H_2SiX_2$ where X= F, Cl, Br, or I (e.g., Reed 1964 discussed $H_2SiF_6$). Bromine and iodine may also be volatilized in elemental form. Some studies apply pyrohydrolysis where samples are only fused with $V_2O_5$ without $Na_2O_2$ or NaOH additions (e.g., Quijano-Rico & Wänke 1969, Dreibus et al. 1979; Menard et al. 2013) and comparison of the results to other studies may suggest that recovery of the halogens in these investigations might not have been fully quantitative ,which seems to be an issue for Br in carbonaceous chondrites. The purpose of the $V_2O_5$ is to obtain better coupling to the RF field of the induction furnace, but $V_2O_5$ will not lead to formation of more volatile halogen-bearing compounds. Fusion with $V_2O_5$ alone may only work well if steam can quantitatively convert halogens into hydrogen halides, and if the intrinsic molar alkali content of the samples exceeds that of the halogen contents (e.g., the atomic CI-chondritic

Na/Cl is around ten). We suggest that fusion with $Na_2O_2$-NaOH in addition to $V_2O_5$, may provide better halogen extraction yields, especially for relatively short steam exposures. The short paper by Easton et al. (1977) describes that pyrohydrolysis yields are too low unless samples were pre-treated. They found up to eight-times larger fluorine concentrations for chemically pre-treated samples of ordinary chondrites than untreated samples with the pyrohydrolysis method. This is described in the fluorine data section 3.1.

Several studies used leaching with water or acid solutions to determine the “soluble” and “non-soluble” fractions in the samples, for example, leaching with hot or cold water was done by Reed & Allen (1966), von Gunten et al. (1965), Tarter (1981), and Sharp et al. (2007). Nakamura et al. (2011) used stepwise leaching with $HNO_3$ and HF for silicate samples for subsequent analyses. Leaching depends on solution temperature, acidity, and duration, which hampers direct comparison from such experiments. The pH of the solutions was generally not reported and other elements dissolved were analyzed only in a few instances so that the type of dissolving salts is not always discernable.

### 2.1.2. Neutron Activation Analysis (NAA)

NAA uses thermal or epithermal neutron capture reactions on halogen nuclei to produce radioactive halogen nuclei which then beta-decay into stable noble gas isotopes. The production of the radioactive halogen isotopes as well as the noble gas isotopes is proportional to the amount of the stable halogens in the samples. The neutron flux and capture geometries in the reactor can be monitored by adding standard samples during the irradiation for calibration so that the halogen concentrations can be calculated quantitatively from the yields of the neutron-capture reaction products. After irradiation and chemical separation from other radioactive nuclei (which is often but not always done to improve counting statistics) measurements proceed to either gamma and/or beta-ray counting of the decay of the radioactive halogens; or to measurements of the decay products, the noble gases.

Many studies used sample irradiation, chemical separation, and counting radiation from decay of the activated halogens. The short half-lives of several activated halogens require fast separation techniques, and incomplete separation is one potential error source (see below).

In both cases quantitative measurements of the activated halogens or the noble gases must be ensured. Potential error sources include the loss of volatile halogens or noble gases during and after irradiation from the samples. Samples can heat up to around 200-400°C while being irradiated in the reactor, e.g., Mittlefheldt (2002) notes that release of volatiles can even rupture the sealed silica tubes used for samples CI and CM samplesduring irradiation which can be prevented by drying the samples before irradiation. This can separate volatile halogen-bearing compounds which are either naturally present in the samples or produced through hot-atom reactions. Further, release of indigenous water can leach out halogens into the irradiation sample container. For example, Krähenbühl et al. (1973b) noted that their low Br results are likely suspect and call for special precautions in handling irradiated samples. Ebihara et al. (1982) carefully sealed their silica vials for irradiation to avoid losses during sample heating, and samples from the reactor were cooled in liquid $N_2$ before opening. They leached vials under oxidizing and reducing conditions to ensure complete recovery of any material that may have condensed onto the walls of the sample vials. Yet they do not rule out that volatile compounds were lost during chemical separation after irradiation, before radiochemical exchange with carrier solutions was completed. Similarly, Mittlefehldt (2002) cautioned his Br results because volatilization loss cannot be excluded. Ozaki & Ebihara (2007) cooled their samples with dry ice during neutron irradiation, and caution that loss of halogens from samples must be prevented. They refer to a study by Takeuchi (1981) who found that filter paper impregnated with bromides and iodides partly lost the halogens (probably as $Br_2$ and $I_2$) to plastic bags in which the paper was stored during irradiation.

Clay et al. (2017) used neutron-irradiation noble gas mass spectrometry (NI-NGMS) and measured the stable noble gas isotopes formed by decay of the radioactive halogens. For quantitative analyses it is necessary to wait for complete decay of the radioactive halogens to the noble gases or to make appropriate corrections for yields depending on half-lives. Clay et al. use laser-heating to drive out the trapped noble gases from the samples after irradiation.

The volatilization of noble gases through recoil and/or heating and desorption during longer irradiation is a well-known issue in the K-Ar and I-Xe dating methods. The iodine-xenon age-dating method uses neutron irradiation to produce $^{128}Xe$ from stable iodine $^{127}I(n, \gamma)^{128}Xe$. Both $^{128}Xe$ and $^{129}Xe$ (from decay of $^{129}I$ (half-life 15.7 my) which was present in the early solar system) can then be measured together to derive the initial $^{129}I/^{127}I$ iodine ratio in the solar system. If the method is calibrated accordingly, the $^{128}Xe$ concentrations can be used to determine the absolute iodine concentration. This is the same principle as used by Clay et al. (2017). Hohenberg (1968), Hohenberg et al. (1981), and Crabb et al. (1982) determined iodine concentrations from $^{128}Xe$ produced in neutron-irradiated samples and found about an order of magnitude lower iodine contents than obtained by other methods such as neutron activation analyses, where the gamma emission of $^{128}I$ during decay to $^{128}Xe$ is monitored (e.g., Goles et al. 1967, Dreibus et al. 1979). Apparently noble gas measurements for $^{128}Xe$ could only quantify iodine in retentive sites. Crabb et al. (1982) refer to work by Hohenberg and coworkers who suspect that some $^{128}Xe$ is lost by recoil during sample irradiation in the reactor, and/or during pre-heating the sample for analyses. Crabb et al. (1982) also note that $^{128}Xe$ could be readily lost from the water-soluble host phase in which a larger proportion of the iodine resides because $^{128}Xe$ did not remain at the original sites of the iodine. Recoil loss is known to be grain-size dependent, as has been shown for argon isotopes (see e.g., Paine et al. 2006, Jourdan et al. 2007) and thus would affect chlorine analyses if noble gases produced from chlorine are analyzed. Since recoil loss depends on the momentum of the recoiling atom, the more massive krypton (from bromine) and xenon (from iodine) might be less affected. Gilmour et al. (2006) claim that losses of $^{128}Xe$ through recoil are not an issue for I-Xe dating since the recoiling nucleus ($^{128}I$) has less than 3eV kinetic energy, but that the activation energy for $^{129}Xe$ (intrinsic to the sample) is above 3.6 eV. Gilmour et al (2006) also claim that recoil effects have never been observed for the I-Xe technique but it is not explained how one would know that there is no effect. Measurements of xenon isotopes from irradiated chondrites are done by stepwise heating and often reveal two or more release peaks, indicating more than one phase is hosting iodine. If $^{128}Xe$ (made from $^{127}I$ during reactor radiation) shows two maxima in gas release, then there are at least two phases that carried the original iodine. The $^{129}Xe$ (intrinsic to the sample and from $^{129}I$ that decayed early in solar system history) should be at the same site(s) if nothing happed to the sample since iodine and xenon entered a condensed phase. Then a correlation of $^{128}Xe$ and $^{129}Xe$ proves that $^{129}Xe$ in the sample came from $^{129}I$ decay, and $^{128}Xe$ is a proxy proportional to the total (stable) iodine content of the sample. Yet, in many instances there is only a good correlation for $^{128}Xe$ with $^{129}Xe$ in the high temperature (above about 1000 C) release fractions, but not very convincing correlations at low temperatures (e.g., Merrihue 1966; Crabb et al. 1982; Hohenberg & Pravdivtseva 2008). Loss of $^{129}Xe$ from less retentive sites is easily possible over the long history of the meteorite parent body (e.g., Gilmour et al. 2006), and for neutron-irradiated iodine in a reactor, loss of $^{128}Xe$ must be prevented in quantitative analyses. The I-Xe dating method relies on the correlation of $^{128}Xe$ and $^{129}Xe$ and using the high temperature release fractions is useful for that. However, for quantitative iodine determinations, all $^{128}Xe$ from iodine in all sites must be retained. If samples contain intrinsic water and have water soluble salts, $^{128}Xe$ from iodine in the salts may not require much energy to be lost from irradiated samples. Mobility of $^{128}Xe$ could be a reason why low temperature correlations for $^{129}Xe$-$^{128}Xe$ are rare in I-Xe dating, however, it must certainly be considered for quantitative iodine analyses.

Clay et al. use the NI-NGMS method and noted that "... Small samples were wrapped in Al foil and, interspersed with monitor minerals, encapsulated under vacuum in fused silica glass tubing before packing in Al canisters for irradiation." Thus, loss from small samples or samples with larger porosity such as CI-chondrites cannot be excluded and it might be necessary to seal samples individually and to break sample containment inside the mass spectrometer.

Clay et al. (2017) use the method described by Ruzie-Hamilton et al. (2016), which is originally from Kendrick (2012) who noted: "The development of solid standards (rather than powdered reference materials) and characterisation of standard reproducibility is an essential step for cross calibration and wider application of microbeam techniques in halogen analysis ...". Kendrick (2012) emphasized the use of "… 0.1–3 mm sized chips of minerals or glasses (not powders)…" and further that "The technique is not easily applied to powdered samples, meaning bulk rock determinations are difficult." The use of powders is not mentioned by Ruzie-Hamilton et al. (2016). Clay et al. describe their procedure: "… samples were either crushed into small chips and sieved, or aliquots extracted from powdered, larger main masses (for example, 1–2-mg-sized aliquots from a sample of up to 1.5 g) where possible …".

With respect to powder and small grain size the problem is compounded for CI-carbonaceous meteorites which are the most porous meteorites. Consolmagno & Britt (1998) and Macke et al. (2011) found a porosity of 35% for the CI-chondrite Orgueil, and porosities for CM and CV chondrites are around 20%. This facilitates loss by heating and potential recoil loss of noble gases from these meteorites.

## 3. Halogen measurements in rock standards and meteorites

In order to assess halogen analyses we compared halogen measurements of rock standards and different meteorite types by authors who also analyzed CI-chondrites. This survey reveals that (1) analyses that are frequently rejected because they are “old” are not any worse than more recent measurements, and (2) analytical problems are endemic for CI-chondrites, due to their mineralogy, i.e., hydrous silicates and several halogen-bearing host phases.

The recommended rock standard concentrations are consensus averages from a large survey of measurements reported in the literature, and recommended values are periodically updated (see references for Gladney &

coworkers quoted in the Tables below). Thus, the results for rock standards from various authors analyzing meteorites also informed the rock-standard concentrations. The recommended concentrations for some rock standards have changed over time. Earlier authors may have found good agreements to recommended rock standard values that validated their analytical techniques whereas comparison to the currently recommended values (listed on the last row for each standard in the tables) may turn out worse. Some groups used rock standards for calibration, and a change in standard values will affect absolute values, but we did not re-calculate any values. In addition, chondritic meteorites often contain silicates, sulfides, metal and/or organic phases whereas standard rocks are usually composed of silicate phases only, and the presence of other phases may affect analytical results. These issues have to be kept in mind when gauging analytical results in the following comparisons.

For each halogen, there are three tables, one for rock standard analyses, one for chondrites (other than CI-chondrites), and one for CI-chondrites. Data included in the Tables are (1) for meteorite falls only to avoid bias from contamination or weathering losses on Earth, and (2) meteorites that were analyzed by authors who also analyzed CI-chondrites. However, this pool of data is extremely small; and we essentially included all available halogen data for meteorite falls that we found in the literature. In all Tables values in square brackets are questionable as noted by the authors, or that values need to be excluded as concluded in this work.

## 3.1. Fluorine

Fluorine data for rock standards, meteorites and CI-chondrites are given in Tables 1-3. The analyses for other chondrites are in Table 2 for assessment of the CI-chondrite data in Table 3. The following discussion shows that except for early data by Reed (1964) and Reed & Jovanovic (1969), analytical results generally agree for terrestrial rocks (as well as basaltic achondrites (eucrites) which are more similar to terrestrial rocks). In stark contrast, chondrite data disagree among different investigators using different analytical methods indicating that there are unique problems with meteorite analyses.

Fluorine analyses by photon activation of $^{19}F(\gamma,n)^{18}F$ were done by Reed (1964), and Reed & Jovanovic (1969, 1971, 1973). The W-1 and G-1 rock standard values (Table 1) by Reed (1964) and Reed & Jovanovic (1969) are much higher than the standard values they measured later (Reed & Jovanovic 1971,1973). Reed (1964) analyzed fluorine in water soluble and non-soluble fractions for several meteorites (Tables 2 & 3). All his values are higher than analyses done by Dreibus et al. (1979) and Allen & Clark (1977). Given the standard and meteorite comparisons, we exclude the high values by Reed (1964), and Reed & Jovanovic (1969) here, because it seems that the early data by Reed (1964) suffered from calibration issues. However, the data by Reed (1964) could be useful for comparing *relative* fluorine concentrations of the meteorites he analyzed, and for estimating the soluble fraction of F from sample leaching experiments. Reed (1964) analyzed two Orgueil samples, each had 64 ppm soluble F, plus 88 or 115 ppm non-soluble F. The soluble amounts appear to be comparable to the pyrohydrolysis results by Dreibus et al. (1979), which is an important fact considered below. Earlier it was thought that the soluble fraction of F in Reed's data could reflect contamination, but this is rather unlikely.

**Table 1. Fluorine in Rock Standards (ppm by mass) also Analyzed in Meteorite Studies, and Literature Survey**

| F, ppm | 1σ | N | Reference |
|---|---|---|---|
| **AGV-1 Andesite Rock Standard** | | | |
| 408 | 30 | 10 | Allen & Clark (1977) |
| 430 | | | Dreibus et al. (1979) |
| 412 | 15 | 2 | Langenauer et al. (1992) |
| [350] | | | Sen Gupta (1968a) |
| 425 | 32 | 35 | Review: Gladney et al. (1992) |
| **BCR-1 Columbia River Basalt Rock Standard** | | | |
| 502 | 50 | 12 | Allen & Clark (1977) |
| 491 | 9 | 1 | Clark et al. (1975) |
| 473 | 23 | 1 | Dreibus et al. (1977, 1979) |
| 491 | 15 | 2 | Langenauer et al. (1992) |
| 450 | 58 | 4 | Sen Gupta (1968a) |
| 500 | | | Sen Gupta (1968b) |
| 490 | 50 | 40 | Review: Gladney et al. (1990) |
| **G-1 Granite Rock Standard** | | | |
| 664 | 30 | 3 | Allen & Clark (1977) |
| 740 | 74 | 10 | Greenland & Lovering (1965) |
| [1077] | 99 | 4 | Reed (1964) |
| [800] | 65 | 1 | Reed & Jovanovic (1969) |
| 705 | 70 | 2 | Shima (1963) |
| 690 | 110 | 34 | Review: Gladney et al. (1991) |
| **G-2 Granite Rock Standard** | | | |
| 1224 | 120 | 12 | Allen & Clark (1977) |
| 1220 | | 4 | Sen Gupta (1968b) |
| 1280 | 80 | 37 | Review: Gladney et al. (1992) |
| **W-1 Diabase Rock Standard** | | | |
| 205 | | 1 | Dreibus et al. (1979) |
| 200 | | 1 | Greenland & Lovering (1965) |
| [489] | 89 | 2 | Reed (1964) |
| [263] | 47 | 2 | Reed & Jovanovic (1969) |
| 187 | | 1 | Reed & Jovanovic (1971) |
| 216 | 11 | 6 | Reed & Jovanovic (1973) |
| 190 | 30 | 1 | Shima (1963) |
| 220 | 30 | 35 | Review: Gladney et al. (1991) |

Note: Values in [ ] are considered uncertain when compared to recommended values, see text. Data by Reed (1964) and Reed & Jovanovic (1969) are systematically high.

Clark et al. (1975), Allen & Clark (1977), and Goldberg et al. (1974) used the proton reaction $^{19}F(p, \alpha\gamma)^{16}O$ and counting high-energy gamma rays. This method does not involve any chemical procedure and should produce reliable F values that do not carry large uncertainties from potentially incomplete chemistry. However, volatilization losses during irradiation cannot be excluded. The rock standard values by Allen & Clark (1977) agree well with recommended standard values. Goldberg et al. note that only small samples can be analyzed and that powdered samples give somewhat smaller values than fragments, e.g., 59-75 ppm for four CM-chondrite powders vs. 53-95 ppm in fragments. The Goldberg et al. data for homogenized samples of Orgueil are 74 ppm and of Ivuna 70 ppm. Unfortunately, Allen & Clark (1977) did not analyze CI-chondrites.

Fisher (1963) used the (n, α) reaction $^{19}F(n, \alpha)^{16}N$ and measurement of the 6.1 MeV gamma ray from the $^{16}N$ decay (half-life 7.4 sec). No rock standard values were reported. Values for meteorites are generally higher (Tables 2,3) than in other studies. He found fluorine contents of 390 and 420 ppm in CI-chondrites. His method requires very large corrections for the bulk oxygen content in the samples because $^{16}N$ is produced by $^{15}N(n, \gamma)^{16}N$, and more importantly, $^{16}O(n, p)^{16}N$. Fisher used the Orgueil oxygen content of 38.3 mass% from Wiik (1956) which, however, is the value stated for volatile-free CI-chondrites, which is much lower than the intrinsic oxygen content of about 46% in CI-chondrites (Lodders 2003). We exclude his CI-chondrite values in Table 3. However, Fisher's F contents for other meteorites (Table 2) were corrected with appropriate oxygen contents and should be useful for comparisons. Sen Gupta (1968a,b) used colorimetry, as did Shima (1963) after sample ion-exchange. Except for AGV-1, Sen Gupta's rock standard values agree within uncertainties, as do Shima's values (Table 1). They did not analyze CI-chondrites but did analyze several other meteorites (Table 2) which should help to compare halogen data obtained by different groups.

**Table 2. Fluorine in Chondrites (Observed Falls)**

| F, ppm | 1σ | N | Meteorite, Year of Fall | Reference; Note |
|---|---|---|---|---|
| **CM Chondrite Falls** | | | | |
| 80 | 5 | 1 | CM2 Essebi 1957 | Goldberg et al. (1974) |
| 59 | 5 | 1 | CM2 Haripura 1921 | Goldberg et al. (1974) |
| [220] | 14 | 2 | CM2 Mighei 1889 | Fisher et al. (1963) |
| 38 | | 1 | CM2 Murchison 1969 | Dreibus et al. (1979) |
| 65 | 5 | 1 | CM2 Murchison 1969 | Goldberg et al. (1974) |
| 37 | | 1 | CM2 Murray 1950 | Dreibus et al. (1979) |
| **CO Chondrite Falls** | | | | |
| 170 | | 1 | CO3 Lance 1872 | Greenland & Lovering (1965) |
| [77] | 18 | 2 | CO3 Lance 1872 | Reed (1964) |
| 30 | | 1 | CO3 Warrenton 1877 | Dreibus et al. (1979) |
| 160 | 5 | 1 | CO3 Warrenton 1877 | Greenland & Lovering (1965) |
| **CV Chondrite Falls** | | | | |
| 58 | 11 | 1 | CV3-ox. Allende 1969 | Allen & Clark 1977 |
| 25 | 12 | 1 | CV3-ox. Allende 1969 | Dreibus et al. (1979) |
| 59 | 5 | 1 | CV3-ox. Allende 1969 | Goldberg et al. (1974) |
| 28 | 1 | 9 | CV3-ox. Allende 1969 | Langenauer & Krähenbühl (1993) |
| 55 | | 1 | CV3-ox. Allende 1969 | Morrison et al. (1987) |
| 170 | 26 | 1 | CV3-red. Mokoia 1908 | Greenland & Lovering (1965) |
| 22 | | 1 | CV3-red. Vigarano 1910 | Dreibus et al. (1979) |
| **CK Chondrite Falls** | | | | |
| 22 | | 1 | CK4 Karronda 1930 | Dreibus et al. (1979) |
| **H Chondrite Falls** | | | | |
| 21 | 6 | 1 | H3.4 Sharps 1921 | Allen & Clark (1977) |
| 81 | 12 | 1 | H4 Forest Vale 1942 | Greenland & Lovering (1965) |
| 8 | | 1 | H4 Kesen 1850 | Dreibus et al. (1979) |
| 69 | 26 | 3 | H4 Kesen 1850 | Fisher (1963) |
| 32 | 10 | 1 | H5 Allegan 1899 | Allen & Clark (1977) |
| 170 | 26 | 1 | H5 Allegan 1899 | Greenland & Lovering (1965) |
| [122] | 29 | 2 | H5 Allegan 1899 | Reed (1964) |
| 39 | 5 | 2 | H5 Ehole 1961 | Shima (1963) |
| 29 | 17 | 1 | H5 Forest City 1890 | Allen & Clark (1977) |
| [89] | 9 | 1 | H5 Miller AR 1930 | Reed (1964); weathered sample |
| 41 | 2 | 1 | H5 Pultusk 1868 | Allen & Clark (1977) |
| [133] | 5 | 1 | H5 Pultusk 1868 | Reed (1964) |
| [162] | 16 | 1 | H5 Richardton 1918 | Reed (1964); weathered sample |
| 40 | 10 | 1 | H5 Sindhri 1901 | Allen & Clark (1977) |
| 17 | | 1 | H6 Charwallas 1834 | Dreibus et al. (1979) |
| 110 | 17 | 1 | H6 Mt. Browne 1902 | Greenland & Lovering (1965) |
| [14] | 2 | 3 | H6 Oakley – find 1895 | Easton et al. (1977); not pretreated |
| 63 | 8 | 3 | H6 Oakley – find 1895 | Easton et al. (1977); pretreated |
| 160 | 24 | 1 | H6 Zhovtnevyi 1938 | Greenland & Lovering (1965) |
| **L Chondrite Falls** | | | | |
| [7] | | 1 | L5 Barwell 1965 | Easton et al. (1977); not pretreated |
| 47 | 34 | 2 | L5 Barwell 1965 | Easton et al. (1977); pretreated |
| 48 | 10 | 1 | L5 Farmington 1890 | Allen & Clark (1977) |
| 250 | 38 | 1 | L5 Farmington 1890 | Greenland & Lovering (1965) |
| 300 | | 1 | L5 Farmington 1890 | Sen Gupta (1968a) |
| 76 | 11 | 1 | L5 Homestead 1875 | Greenland & Lovering (1965) |
| [4] | | 1 | L5 Ohuma 1963 | Easton et al. (1977); not pretreated |
| 33 | 1 | 2 | L5 Ohuma 1963 | Easton et al. (1977); pretreated |
| 9 | | 1 | L6 Alfianello 1883 | Dreibus et al. 1979) |
| 32 | 1 | 1 | L6 Bruderheim 1960 | Allen & Clark (1977) |
| 11 | 1 | 1 | L6 Bruderheim 1960 | Dreibus et al. (1979) |
| [123] | 23 | 4 | L6 Bruderheim 1960 | Reed (1964) |
| 52 | 1 | 1 | L6 Canakkale 1964 | Allen & Clark (1977) |
| 42 | 12 | 1 | L6 Colby WI 1917 | Allen & Clark (1977) |
| 44 | 15 | 1 | L6 Forksville 1924 | Allen & Clark (1977) |
| [156] | 67 | 3 | L6 Harleton 1961 | Reed (1964) |
| [43]** | 16 | 1 | L6 Holbrook 1912 | Allen & Clark (1977) |
| [189]** | | 1 | L6 Holbrook 1912 | Fisher (1963) |
| [130]** | | 1 | L6 Holbrook 1912 | Greenland & Lovering (1965) |
| [16]** | 1 | 1 | L6 Holbrook 1912 | Langenauer & Krähenbühl (1993) |
| [149] | 13 | 1 | L6 Kunashak 1949 dark | Reed (1964) |
| [145] | 21 | 1 | L6 Kunashak 1949 light | Reed (1964) |
| 18 | | 1 | L6 Leedey 1943 | Dreibus et al. (1979) |
| 11 | | 1 | L6 Mocs 1860 | Dreibus et al. (1977, 1979) |
| [10] | | 1 | L6 Mocs 1860 | Easton et al. (1977); not pretreated |
| 86 | | 1 | L6 Mocs 1860 | Easton et al. (1977); pretreated |
| 147 | 36 | 4 | L6 Mocs 1860 | Fisher (1963) |
| 160 | | 1 | L6 Mocs 1860 | Greenland & Lovering (1965) |
| [119] | 13 | 2 | L6 Mocs 1860 | Reed (1964) |
| 190 | 29 | 1 | L6 Narellan 1928 | Greenland & Lovering (1965) |
| [122] | 5 | 3 | L6 New Concord 1860 | Reed (1964) |
| 32 | 5 | 1 | L6 Peace River 1963 | Allen & Clark (1977) |
| 118 | 18 | 1 | L6 Perpeti 1935 | Greenland & Lovering (1965) |
| 35 | 15 | 1 | L6/7 Shaw 1937 | Allen & Clark (1977) |
| 120 | 18 | 1 | L6 St. Michel 1910 | Greenland & Lovering (1965) |
| **LL Chondrite Falls** | | | | |
| 8 | | 1 | L/LL4 Bjurböle 1899 | Dreibus et al. (1979) |
| 100 | 15 | 1 | L/LL4 Bjurböle 1899 | Greenland & Lovering (1965) |
| 68 | 10 | 1 | L3.6 Khohar 1910 | Greenland & Lovering (1965) |
| 49 | 22 | 1 | LL4 Soko-Banja 1877 | Allen & Clark (1977) |
| 160 | 24 | 1 | LL5 Olivenza 1924 | Greenland & Lovering (1965) |
| 63 | 4 | 2 | LL6 Benton 1949 | Sen Gupta (1968a) |
| 180 | 27 | 1 | LL6 Mangwendi 1934 | Greenland & Lovering (1965) |
| **EH Chondrite Falls** | | | | |
| 64 | | 1 | EH4 Abee 1952 | Dreibus et al. (1979) |
| 280 | 42 | 1 | EH4 Abee 1952 | Greenland & Lovering (1965) |
| [228] | 16 | 2 | EH4 Abee 1952 | Reed (1964) |
| 275 | 35 | 2 | EH4 Abee 1952 | Sen Gupta (1968a) |
| 137 | 16 | 4 | EH4 Indarch 1891 | Fisher (1963) |
| 220 | 33 | 1 | EH4 Indarch 1891 | Greenland & Lovering (1965) |
| [19.1] | 8.6 | 1 | EH4 Indarch 1891 | Reed (1964); hot NaF soln. |
| [246] | 19 | 1 | EH4 Indarch 1891 | Reed (1964) |
| 140 | 21 | 1 | EH5 St. Marks 1903 | Greenland & Lovering (1965) |
| **EL Chondrite Falls** | | | | |
| 250 | | 1 | EL6 Hvittis 1901 | Greenland & Lovering (1965) |
| [134] | 24 | 3 | EL6 Hvittis 1901 | Reed (1964) |
| 180 | | 1 | EL6 Khairpur 1873 | Greenland & Lovering (1965) |
| 54 | 5 | 1 | EL6 Neuschwanstein 2002 | Zipfel et al. (2010) |
| 100 | | 1 | EL6 Pillistfer 1863 | Greenland & Lovering (1965) |

Note: Mainly chondrite falls. Data in square brackets were excluded for averages for reasons described in the text. **Holbrook 1912: Samples were collected for 50+ years after fall, weathering were noted for some specimens in the references.

Greenland & Lovering (1965) employed a spectrographical method using CaF band emission after sample volatilization and selective element capture. Their rock standard value for “W-1” (Table 1) compares well to the current recommended value by Gladney et al. (1991). They found a higher value for standard “G-1”; but agree within given uncertainties. Greenland & Lovering (1965) compared their results to wet-chemical analyses of various rock-types and agreement was quite reasonable. Dreibus et al. (1979) note that their values are generally lower than those by Reed (1964) and Greenland & Lovering (1965) (see Table 2) but provide no compelling arguments other than that they “performed numerous experiments to test the reliability of our fluorine analysis and we are confident on the accuracy of our results.” The meteorite data by Greenland & Lovering (1965) are often as large as or larger than those by Reed (1964) whose values we excluded because his values for the rock standards are too high. However, there are no obvious reasons to exclude the values by Greenland & Lovering (1965) from consideration other than that one of their rock standard measurements is too high. In Table 2, we summarized results for fluorine in different chondrites. If Reed’s (1964) values are too high, then the even higher values by Greenland & Lovering (1965) for some meteorites (e.g., Lance, Allegan, Hvittis, Mocs) could be suspect.

The data by Dreibus et al. (1979) and Wänke et al. (1977) by pyrohydrolysis and ion-selective electrode agree well with recommended rock standard values (Table 1). Their average F concentration for Orgueil from four measurements is 58±9 ppm. Menard et al. (2013) found 100 ppm F for Orgueil from pyrohydrolysis and ion chromatography; they have no samples in common with those listed in Tables 1 and 2 that can be compared. Still, the F concentrations by Dreibus et al (1979) for other meteorites are often lower than found in other studies. Langenauer & Krähenbühl (1993) used the same method as Dreibus et al. (1979) and their results are very similar. However, Easton et al. (1977) have shown that a pre-hydrolysis treatment of samples leads to higher extraction yields of fluorine from ordinary chondrites. Without the pre-treatment, Easton et al. found 10 ppm F for Mocs and essentially reproduced the 11 ppm that Dreibus et al. (1977,1979) found for this meteorite (see Table 2). However, with sample pre-treatment Easton et al. (1977) found 86 ppm for Mocs. Easton et al. (1977) did not have other meteorite analyses in common with other studies, and they only analyzed ordinary chondrites. Although the F concentrations measured by Reed (1964) could be too high, the pyrohydrolysis results are close to the NaF-soluble fraction in the Orgueil meteorite determined by Reed (1964). All this indicates that soluble fluorine (from salts) is easily mobilized in pyrohydrolysis treatment, but that some portion of the fluorine could be more tightly bound in these meteorites and is not captured by pyrohydrolysis for quantitative analyses (see section 2).

The average F concentrations for Orgueil determined by Goldberg (1974), Dreibus et al. (1979), Menard et al. (2013) and Greenland & Lovering (1965) is 105±60 ppm. Adding the Reed data gives 115±57. We select the average of 105±60 ppm F for Orgueil from the four studies. From two studies of Ivuna, we obtain 67±5 ppm F (Table 3). There are no F data for other CI-chondrites.

**Table 3. Fluorine in CI-Chondrite Falls**

| F, ppm | 1σ | N | Reference; Note |
|---|---|---|---|
| | | | **Ivuna 1939** |
| 63 | 6 | 1 | Dreibus et al. (1979) |
| 70 | 5 | 1 | Goldberg et al. (1974) |
| **67** | **5** | **2** | **Average Ivuna** |
| | | | **Orgueil 1864** |
| 54 | 2 | 4 | Dreibus et al. (1979) |
| [405] | 21 | 1 | Fisher (1963) |
| 74 | 5 | 1 | Goldberg et al. (1974) |
| 190 | 29 | 1 | Greenland & Lovering (1965) |
| 100 | | 1 | Menard et al. (2013) |
| [62] | 6 | 2 | Reed (1964); NaF soluble only |
| [158] | 19 | 2 | Reed (1964) |
| **105** | **60** | **4** | **Average Orgueil** |

Note: N = number of individual analyses. Values in [ ] excluded from averages, see text.

We did not average over the 14 individual analyses from all CI-chondrites to avoid giving one investigation too much statistical weight just because more measurements were made. The straight average for the two CI-chondrite averages is 86±20 ppm with a 1-sigma standard deviation. The average weighted by the number of studies is 92±20 ppm, which we adopt as representative CI-chondrite value here. The F concentration corresponds to 1270(±270) per $10^6$ Si atoms (see also sections 3.6 and 4.1).

## 3.2. Chlorine, Bromine, and Iodine

Combined chlorine, bromine and iodine analyses by neutron activation were done by Reed & Allen (1966), Reed & Jovanovic (1969), Goles et al. (1967), and Dreibus et al. (1979). Several other studies used RNAA or related methods to obtain halogen data; recently Ebihara & Sekimoto (2019) reported results for CI-chondrites that are not included in the data base here since they were not listed in their abstract. The Cl, Br, and I determinations for the CI-chondrite Orgueil by Clay et al. (2017) are systematically low when compared to other studies on meteorites. Clay et al. (2017) do not report standard rock measurements but we can compare their measurements to other meteorite measurements. However, there is generally little overlap for analyses of meteorite falls (Clay et al. erroneously list Barratta, Clovis, and Efremovka as observed meteorite falls). Ebihara & Sekimoto (2019) reported new Cl, Br, and I data for CI-chondrites at the Lunar and Planetary Science Conference. Their results confirm our current assessment and recommendations.

### *3.2.1. Chlorine*

Rock standard values for chlorine are shown in Table 4. Note that Gladney et al. (1991) have two different recommendations for the consensus value of the “G-1” standard in their tables for unknown reasons. Chlorine analyzed by RNAA by Dreibus et al. (1979) agrees well for all standards in common. Reed & Allen (1966) and Reed & Jovanovic (1969) found systematically lower chlorine in their early studies than the recommended values, and generally lower values for meteorites (Table 5), which is why their data are not considered further.

Greenland & Lovering (1965) have no chlorine measurements for rock standards in common with other studies but the meteorite comparisons are instructive. Their chlorine values for chondrites are typically somewhat higher than those by others (Table 5) except for the CI-chondrite Orgueil, which is significantly lower at 290 ppm (Table 6). Goles et al. (1967) note that the Orgueil datum by Greenland & Lovering (1965) is low because of ineffective capture of volatile halogen-bearing organics during cold-trapping of the volatilized sample (see section 2).

**Table 4. Chlorine in Rock Standards (ppm by mass) also Analyzed in Meteorite Studies, and Literature Survey**

| Cl, ppm | 1σ | N | Reference |
|---|---|---|---|
| **AGV-1 Andesite Rock Standard** | | | |
| 118 | | 1 | Dreibus et al. (1979) |
| 156 | | 1 | Fujitani & Nakamura (2006) |
| 118 | 7 | 2 | Langenauer et al. (1992) |
| 138 | | 1 | Nakamura et al. (2011) |
| 121 | | 1 | Tarter (1981) |
| 126 | 27 | 25 | Review: Gladney et al.(1992) |
| **BCR-1 Columbia River Basalt Rock Standard** | | | |
| 43.2 | 2.5 | 1 | Clark et al. (1975) |
| 50.5 | 2.5 | 1 | Dreibus et al. (1977, 1979) |
| 55 | | 1 | Jovanovic & Reed (1975) |
| 50 | 4 | 2 | Langenauer et al. (1992) |
| 68 | | 4 | Tarter (1981) |
| 59 | 8 | 26 | Review: Gladney et al. (1990) |
| **G-1 Granite Rock Standard** | | | |
| 19 | 0.2 | 1 | Reed & Allen (1966) |
| 40 | 18 | 3 | Review Gladney et al. (1991), Table 2 consensus value |
| 53 | 20 | 10 | Review: Gladney et al. (1991); their Tab. 4 consensus value |
| **G-2 Granite Rock Standard** | | | |
| 48 | | 1 | Tarter (1981) |
| 69 | 24 | 24 | Review: Gladney et al. (1992) |
| **W-1 Diabase Rock Standard** | | | |
| 203 | | 1 | Dreibus et al. (1979) |
| 212 | | 1 | Magenheim et al. (1994) |
| [156] | | 4 | Quijano-Rico & Wänke (1969) |
| [25.9] | 0.6 | 2 | Reed & Allen (1966) |
| [16.3] | 1.3 | 1 | Reed & Jovanovic (1969) |
| 205 | 35 | 16 | Review: Gladney et al. (1991) |

Note. N = number of individual analyses. Values in [ ] are considered uncertain when compared to recommended values, see text.

The chlorine values for CI-chondrites by Clay et al. (2017) are very low (Table 6), whereas their values for Murchison (CM2) and Murray (CM2) compare better with other results (Table 5). Their value of 85 ppm for Allende (CV3) is much lower than in any other study, and at the opposite end of the 330 ppm structurally-bound chlorine reported by Sharp et al. (2013). The Allende meteorite has been studied extensively and it would be hard to believe that all other analyses for it in Table 5 are incorrect (The Allende reference sample value is 291±36 ppm, Jarosewich et al. 1987). The only other chondrite falls to which we can compare chlorine data from Clay et al. (2017) are enstatite chondrites, and the data by Clay et al. are much lower than other values. As in other chondrites, sample homogeneity is a problem when only a few milligrams are used for analyses. Overall, the very low values for chlorine in CI-chondrites by Clay et al. seem to be an artifact because (1) their analysis method is unsuitable for the water-bearing CI-chondrites because samples with large intrinsic water content enable halogen leaching in the reactor during irradiation, (2) the small samples analyzed are not representative for the whole rock, (3) the naturally fine-grained powdery CI-chondrites facilitate recoil-loss, and (4) losses of volatile halogen-bearing salts and organics during sample heating cannot be excluded (see section 2).

**Table 5 Chlorine in Chondrites (Falls Only) ***

| Cl, ppm | 1σ | N | Class, Meteorite, Year of Fall | Reference; Note |
|---|---|---|---|---|
| **CM Chondrite Falls** | | | | |
| 470 | 57 | 2 | CM2 Mighei 1889 | Goles & Anders (1962) |
| [350] | 10 | 1 | CM2 Mighei 1889 | Reed & Allen (1966) |
| [231] | 9 | 2 | CM2 Murchison 1969 | Clay et al. (2017) |
| 180 | | 1 | CM2 Murchison 1969 | Dreibus et al. (1979) |
| 242 | 7 | 2 | CM2 Murchison 1969 | Magenheim et al. (1994) |
| [739] | 5 | 1 | CM2 Murchison 1969 | Nakamura et al. (2011) who note suspiciously high value. |
| 179 | 10 | 8 | CM2 Murchison 1969 | Tarter (1981) |
| [151]** | 9 | 2 | CM2 Murray 1950 | Clay et al. (2017) |
| 186** | | | CM2 Murray 1950 | Dreibus et al. (1979) |
| 200** | 14 | 2 | CM2 Murray 1950 | Goles et al. (1967) |
| [108]** | | | CM2 Murray 1950 | Quijano-Rico & Wänke (1969) who indicated preliminary value. |
| 310** | | | CM2 Murray 1950 | Sharp et al. (2013) |
| 368** | 46 | 6 | CM2 Murray 1950 | Tarter (1981) |
| **CO Chondrite Falls** | | | | |
| 270 | | 1 | CO3 Felix 1900 | Goles et al. (1967) |
| 320 | | 1 | CO3 Kainsaz 1937 | Sharp et al. (2013) |
| [730] | | 1 | CO3 Lance 1872 | Daubrée (1872) |
| 274 | | 2 | CO3 Lance 1872 | Goles et al. (1967) |
| 350 | 29 | 1 | CO3 Lance 1872 | Greenland & Lovering (1965) |
| 248 | | 1 | CO3 Lance 1872 | Quijano-Rico & Wänke (1969) |
| [125] | 3 | 1 | CO3 Lance 1872 | Reed & Allen (1966) |
| 290 | | 1 | CO3 Ornans 1868 | Sharp et al. (2013) |
| 226 | | 1 | CO3 Warrenton 1877 | Dreibus et al. (1979) |
| 360 | 30 | 1 | CO3 Warrenton 1877 | Greenland & Lovering (1965) |
| **CV Chondrite Falls** | | | | |
| 220 | 15 | 2 | CV3-ox. Allende 1969 | Bonifacie et al. (2007) |
| [85] | 3 | 1 | CV3-ox. Allende 1969 | Clay et al. (2017) |
| 237 | 12 | 1 | CV3-ox. Allende 1969 | Dreibus et al. (1979) |
| 320 | 30 | 1 | CV3-ox. Allende 1969 | Ebihara et al. (1986) |
| 260 | 100 | 1 | CV3-ox. Allende 1969 | Izawa et al. (2010) |
| [400] | | 1 | CV3-ox. Allende 1969 | Kato et al. (2000) who note suspic. high value. |
| 228 | 8 | 6 | CV3-ox. Allende 1969 | Langenauer & Krähenbühl (1993) |
| 223 | 18 | 1 | CV3-ox. Allende 1969 | Magenheim et al. (1995) |
| 220 | | 1 | CV3-ox. Allende 1969 | Mason (1975) |
| 336 | 22 | 2 | CV3-ox. Allende 1969 | Nakamura et al. (2011) |
| 372 | 9 | 5 | CV3-ox. Allende 1969 | Oura et al. (2002) |
| 291 | 24 | 6 | CV3-ox. Allende 1969 | Ozaki & Ebihara (2007) |
| 330 | | | CV3-ox. Allende 1969 | Sharp et al. (2013) who note structurally bound Cl only |
| 236 | 24 | 23 | CV3-ox. Allende 1969 | Tarter (1981) |
| 256 | | 1 | CV3-ox Bali 1907 | Quijano-Rico & Wänke (1969) |
| 423 | | 1 | CV3-ox Grosnaja 1861 | Quijano-Rico & Wänke (1969) |
| 370 | 30 | 1 | CV3-red. Mokoia 1908 | Greenland & Lovering (1965) |
| 302 | 18 | 2 | CV3-red. Mokoia 1908 | Tarter (1981) |
| 149 | | 1 | CV3-red. Vigarano 1910 | Dreibus et al. (1979) |
| 270 | | 1 | CV3-red. Vigarano 1910 | Sharp et al. (2013); weathered sample collected 1 months after fall, authors note questionable value, may contain terrestrial contamination in addition to inherent water soluble chlorine |
| **CK Chondrite Falls** | | | | |
| 215 | | 1 | CK4 Karoonda | Dreibus et al. (1979) |
| 310 | | 1 | CK4 Karoonda | Greenland & Lovering (1965) |
| [45] | | | CK4 Karoonda | Quijano-Rico & Wänke (1969) who indicatepreliminary value. |
| [117] | | | CK4 Karoonda | Reed & Allen (1966) |
| 257 | | 1 | CK4 Kobe 1999 | Nakamura et al. (2011) |
| 261 | 12 | 4 | CK4 Kobe 1999 | Oura et al. (2002) |
| **H Chondrite Falls** | | | | |
| 140 | 42 | 2 | H3-6 Zag 1998 | Sharp et al. (2013) |
| 90 | | 1 | H3.8 Dhajala 1876 | Sharp et al. (2013) |
| 76 | 15 | 2 | H4 Forest Vale 1942 | Clay et al. (2017) |
| 170 | | 1 | H4 Forest Vale 1942 | Greenland & Lovering (1965) |
| 90 | | 1 | H4 Kesen 1850 | Dreibus et al. (1979) |
| 72 | | 1 | H4 Kesen 1850 | Quijano-Rico & Wänke (1969) |
| 104 | 11 | 1 | H4 Kesen 1850 | Takaoka et al. (1989)90 |
| 148 | | 1 | H4 Monroe 1849 | Garrison et al. (2000) |
| 107 | 8 | 3 | H4 Monroe 1849 | Tarter (1981) |
| 57 | | 1 | H5 Allegan 1899 | Greenland & Lovering (1965) |
| 7 | | 1 | H5 Allegan 1899 | Quijano-Rico & Wänke (1969) |
| [9.1] | 1.6 | 2 | H5 Allegan 1899 | Reed & Allen (1966) |
| 11 | 3 | 4 | H5 Allegan 1899 | Tarter (1981) |
| 52 | | 1 | H5 Pantar (light) 1938 | Goles et al. (1967) |
| [50] | 3 | 1 | H5 Pantar (light) 1938 | Reed & Allen (1966) |
| 56 | | 1 | H5 Pantar (light) 1938 | Quijano-Rico & Wänke (1969) |
| **L Chondrite Falls** | | | | |
| 68** | | 1 | L/LL4 Bjurböle 1899 | Dreibus et al. (1979) |
| 92** | | 1 | L/LL4 Bjurböle 1899 | Greenland & Lovering (1965) |
| 88** | | 1 | L/LL4 Bjurböle 1899 | Hohenberg et al. (1981) |
| 84** | | 1 | L/LL4 Bjurböle 1899 | Quijano-Rico & Wänke (1969) |
| 120** | | 1 | L/LL4 Bjurböle 1899 | Sharp et al. (2013) note questionable value, may contain terrestrial contamination in addition to inherent water soluble chlorine |
| 99** | 16 | 3 | L/LL4 Bjurböle 1899 | Tarter (1981) |
| 270 | | 1 | L3.6 Khohar 1910 | Greenland & Lovering (1965) |
| ≥ 445 | | | L3.6 Khohar 1910 | Muenow et al. (1995) note release starts at 950C, not all Cl released by 1300C, lower limit. |
| 81 | 4 | 3 | L4 Saratov 1918 | Tarter (1981) |
| 110 | | 1 | L4 Saratov 1918 | Vinogradov et al. (1960) |
| 140 | | 1 | L5 Homestead 1875 | Greenland & Lovering (1965) |
| 243 | | 1 | L5 Homestead 1875 | Tarter (1981) |
| 132 | | 1 | L6 Alfianello 1883 | Dreibus et al. (1979) |
| 208 | | 1 | L6 Alfianello 1883 | Tarter (1981) |
| 100** | | 1 | L6 Bruderheim 1960 | Dreibus et al. (1979) |
| 137** | 14 | | L6 Bruderheim 1960 | Ebihara et al. (1997) |
| 67** | | 1 | L6 Bruderheim 1960 | Garrison et al. (2000) |
| 101** | 22 | 6 | L6 Bruderheim 1960 | Goles et al. (1967) |
| 91** | 12 | 1 | L6 Bruderheim 1960 | Kato et al. (2000) |
| [3.3] ** | 1.1 | 5 | L6 Bruderheim 1960 | Reed & Allen (1966), 1x 59 ppm omitted |
| 71** | 17 | 9 | L6 Bruderheim 1960 | Tarter (1981); 1x 221 ppm omitted |
| 103** | 12 | 3 | L6 Bruderheim 1960 | Von Gunten et al. (1965) |
| 105 | | 1 | L6 Leedey 1943 | Begemann (1964) |
| 112 | | 1 | L6 Leedey 1943 | Dreibus et al. (1979) |
| 93 | 52 | 5 | L6 Leedey 1943 | Garrison et al. (2000) |
| 105 | 81 | 10 | L6 Leedey 1943 | Tarter (1981) |
| 111 | 11 | 3 | L6 Leedey 1943 | Von Gunten et al. (1965) |
| 110 | | 1 | L6 Mocs 1860 | Behne (1953) |
| 95 | | | L6 Mocs 1860 | Dreibus et al. (1979) |
| 70.5 | | 2 | L6 Mocs 1860 | Goles et al. (1967) |
| 220 | | 1 | L6 Mocs 1860 | Greenland & Lovering (1965) |
| 136 | 52 | 5 | L6 Mocs 1860 | Von Gunten et al. (1965) |
| **LL Chondrite Falls** | | | | |
| 131 | | 1 | LL3.6 Parnallee 1857 | Quijano-Rico & Wänke (1969) |
| 200 | | 1 | LL3.6 Parnallee 1857 | Sharp et al. (2013) |
| 130 | 18 | 3 | LL3.6 Parnallee 1857 | Tarter (1981) |
| 230 | | 1 | LL5 Olivenza 1924 | Greenland & Lovering (1965) |
| 37 | 7 | 3 | LL5 Olivenza 1924 | Tarter (1981) |
| **EH Chondrite Falls** | | | | |
| 243 | | 1 | EH3 Parsa 1942 | Garrison et al. (2000) |
| 590 | | 3 | EH3 Qingzhen 1976 | Grossman et al. (1985) |
| 1000 | | 1 | EH3 Qingzhen 1976 | Sharp et al. (2013) |
| 560 | | 1 | EH4 Abee 1952 | Dreibus et al. (1979) |
| 587 | 74 | 3 | EH4 Abee 1952 | Garrison et al. (2000) |
| 750 | 90 | 1 | EH4 Abee 1952 | Goles et al. (1967) |
| 994 | | | EH4 Abee 1952 | Quijano-Rico & Wänke (1969) |
| [432] | 10 | 1 | EH4 Abee 1952 | Reed & Allen (1966) |
| 740 | | | EH4 Abee 1952 | Sharp et al. (2013) |
| 524 | 34 | 4 | EH4 Abee 1952 | Tarter (1981) |
| 530 | | 1 | EH4 Abee 1952 | Von Gunten et al. (1965) |
| [84] | 6 | 3 | EH4 Indarch 1891 | Clay et al. (2017) |
| 400 | | 1 | EH4 Indarch 1891 | Garrison et al. (2000) |
| 570 | | 1 | EH4 Indarch 1891 | Goles et al. (1967) |
| 903 | 25 | 2 | EH4 Indarch 1891 | Greenland & Lovering (1965) |
| [675] | 134 | 2 | EH4 Indarch 1891 | Reed & Allen (1966) |
| 825 | 39 | 6 | EH4 Indarch 1891 | Tarter et al. (1980) |
| [51] | 3 | 1 | EH5 St. Marks 1903 | Clay et al. (2017) |
| 97 | | 1 | EH5 St. Marks 1903 | Garrison et al. (2000) |
| 210 | | 1 | EH5 St. Marks 1903 | Greenland & Lovering (1965) |
| 170 | 22 | 6 | EH5 St. Marks 1903 | Tarter (1981) |
| **EL Chondrite Falls** | | | | |
| [14] | 1 | 2 | EL6 Daniel's Kuil 1868 | Clay et al. (2017) |
| 210 | | 1 | EL6 Eagle | Sharp et al. (2013) |
| 297 | | 1 | EL6 Hvittis 1901 | Garrison et al. (2000) |
| 210, [78] | | 2 | EL6 Hvittis 1901 | Goles et al. (1967) |
| 250 | | 1 | EL6 Hvittis 1901 | Greenland & Lovering (1965) |
| 234 | | 1 | EL6 Hvittis 1901 | Quijano-Rico & Wänke (1969) |
| [323] | | | EL6 Hvittis 1901 | Reed & Allen (1966) |
| 222 | 31 | 2 | EL6 Hvittis 1901 | Von Gunten et al. (1965) |
| 113 | | 1 | EL6 Khaipur 1873 | Garrison et al. (2000) |
| 230 | | 1 | EL6 Khaipur 1873 | Greenland & Lovering (1965) |
| 338 | 17 | 1 | EL6 Neuschwanstein 2002 | Zipfel et al. (2010) |
| 70 | | 1 | EL6 Pillistfer 1863 | Garrison et al. (2000) |
| 160 | 15 | 1 | EL6 Pillistfer 1863 | Greenland & Lovering (1965) |

Note. N = number of individual analyses. Values in [ ] excluded from averages, see text
* Only meteorite falls.
** Notes for some meteorites: Bjurböle fell into frozen lake. Bruderheim is known to be heterogeneous.
Murray is known to have lower volatiles (e.g., alkalies) than other CM; some samples only collected sometime after fall

Chlorine concentrations agree well in larger (> 50 mg) aliquots of larger homogenized samples in the CI-chondrite Orgueil (Table 6) analyzed with nuclear radiation techniques (e.g., Dreibus et al. 1979, Goles et al. 1967, Islam et al. 2011, Palme 2019). The values by Dreibus et al. (1979) with chemical separation agree within uncertainties to data from ion-chromatography (Menard et al. 2013, Magenheim et al. 1994, and Tarter 1980). There are larger variations among the three studies using the latter technique: Sharp et al. (2013) give a total of 970 ppm Cl for Orgueil in their text and from the structurally bound Cl given as 630 ppm and the water soluble of 340 ppm that could contain contamination. In their Table 3, Sharp et al. (2013) indicate a total of 950 ppm for Orgueil which could be questionable because of terrestrial contamination in addition to inherent water soluble chlorine. Assuming that 340 ppm (35% of total) is contamination, their lower limit of "true" Cl is 970-340 = 630 ppm.

We have included the two wet-chemical analyses by Pisani (1864) and Cloëz (1864) that were done within weeks after the Orgueil meteorite fall. Chlorine can be reliably determined quantitatively by gravimetric methods and there is no obvious reason to exclude these results. The chlorine in Ivuna by Sharp et al. (2007) was described as a partial analysis and is therefore excluded. Palme & Zipfel (2021) has three Cl determinations for Ivuna, but we excluded one analysis with 35% uncertainty in his average of 663 ppm listed for Ivuna in Table 6. There is only one analyses for chlorine in Tonk with 25% uncertainty, which we did not include in the group mean. The weighted average (by number of studies per meteorite) from Alais, Ivuna and Orgueil gives 717±110 ppm Cl where the uncertainty of the Orgueil averages is adopted to reflect the wide range in concentrations. This corresponds to 5270(±810) Cl atoms per $10^6$ Si.

**Table 6 Chlorine in CI-Chondrite Falls**

| Cl, ppm | 1σ | N | Reference; Note |
|---|---|---|---|
| | | | **Alais 1806** |
| [840] | 576 | 30 | Greshake et al. 1998; 50-micron-size samples |
| 710 | | 1 | Palme & Zipfel (2021) |
| **710** | | **1** | **Average Alais** |
| | | | **Ivuna 1939** |
| 698 | 35 | 1 | Burnett et al. (1989) |
| 750 | | 1 | Dreibus et al. (1979) |
| 705 | 112 | 2 | Goles et al. (1967) |
| 663 | 39 | 3 | Palme & Zipfel (2021) |
| [278] | 53 | 2 | Reed & Allen (1966) |
| [290] | | | Sharp et al. (2007); partial analysis |
| **704** | **36** | **4** | **Average Ivuna** [a] |
| | | | **Orgueil 1864** |
| [115] | 36 | 2 | Clay et al. (2017) |
| 730 | | | Cloëz (1864) |
| 678 | 122 | 4 | Dreibus et al. (1979) |
| 780 | 60 | 3 | Goles et al. (1967) |
| [290] | | 1 | Greenland & Lovering (1965) |
| [550] | 310 | 31 | Greshake et al. (1998); grand average of 50-micrometer samples |
| 720 | | 1 | Grossman et al. (1985) |
| 780 | 2 | 2 | Islam et al. (2011) |
| 679 | | 1 | Magenheim et al. (1994) |
| 730 | | 1 | Menard et al. (2013) |
| 600 | 74 | 7 | Palme & Zipfel (2021) |
| 760 | | 1 | Pisani (1864) |
| [210) | | 1 | Reed & Allen (1966) |
| 950 | | 1 | Sharp et al. (2013) |
| 536 | 142 | 2 | Tarter et al. (1980) |
| **722** | **107** | **11** | **Average Orgueil** [a] |
| | | | **Tonk 1911** |
| 510 | 130 | 1 | Palme & Zipfel (2021) |

Note. N = number of individual analyses.Values in [ ] excluded from averages, see text. Reed & Allen (1966) are systematically low.

### 3.2.2. Bromine

Many more data for bromine are available than for chlorine. Bromine concentrations can be obtained routinely with instrumental neutron activation analysis (INAA) which requires no chemistry and provides reliable data for uncontaminated meteorite samples, provided losses are minimized from samples during irradiation (see above). Statistical uncertainties of Br analyses in bulk chondrites with INAA are often around 10% or below (Palme 2019). Results of Br analyses for rock standards are in Table 7, for chondrites in Table 8, and Table 9 gives available data for CI-chondrites. Rock standard values for Br in G-1 and W-1 by Reed & Allen (1966) and Reed & Jovanovic (1969) agree with recommended values within uncertainties. Their value for BCR-1 from a later study (Jovanovic & Reed 1975) is too high in comparison to others (Table 7). Chondrite values by Reed & Allen (1966) often fall toward the higher end of bromine values reported by others (Table 8). However, their Br values for CI-chondrites are at the low end of values (Table 9), just opposite to the trends seen comparing other chondrite groups. This behavior is also apparent in some of the other studies that use NAA; the CI- and CM chondrite results for Br are generally quite variable (Tables 8,9). Bromine values determined by Dreibus et al. (1979) for rock standards (Table 7) and ordinary chondrites (Table 8) agree, but their values for carbonaceous chondrites are systematically at the lower end. This is another indication that carbonaceous chondrites are inherently difficult to analyze with NAA and subsequent chemical assay and/or pyrohydrolysis.

**Table 7. Bromine in Rock Standards (ppm by mass) also Analyzed in Meteorite Studies, and Literature Survey**

| Br, ppm | 1σ | N | Reference |
|---|---|---|---|
| **AGV-1 Andesite Rock Standard** | | | |
| 0.345 | | 1 | Dreibus et al. (1979) |
| 0.380 | 0.05 | 1 | Langenauer et al. 1992 |
| [0.560] | | 1 | Lieberman (1966) |
| 0.32 | 0.17 | 7 | Review: Gladney et al.1992 |
| **BCR-1 Columbia River Basalt Rock Standard** | | | |
| 0.046 | 0.002 | 2 | Anders et al. 1971 |
| 0.077 | 0.004 | | Dreibus et al. (1979) |
| 0.0558 | | | Gros et al. (1976) |
| 0.056 | 0.003 | | Hertogen et al. (1977 ) |
| [0.152] | 0.023 | 2 | Jovanovic & Reed (1975) |
| 0.079 | 0.030 | 17 | Keays et al. (1974) |
| 0.060 | 0.033 | 5 | Krähenbühl et al. (1973a) |
| 0.092 | 0.02 | | Krähenbühl et al. (1973b) |
| 0.070 | 0.02 | 1 | Langenauer et al. (1992) |
| 0.060 | | | Laul et al. (1972a) |
| 0.093 | 0.007 | 2 | Laul et al. (1972b) |
| [0.18] | | 1 | Lieberman (1966) |
| 0.064 | | 1 | Treiman et al. (1986) |
| 0.072 | 0.025 | 27 | Review: Gladney et al. (1990) |
| **G-1 Granite Rock Standard** | | | |
| 0.13 | 0.04 | 1 | Reed & Allen (1966) |
| 0.140 | 0.026 | 3 | Review Gladney et al. (1991) |
| **W-1 Diabase Rock Standard** | | | |
| 0.21 | | 1 | Dreibus et al. (1979) |
| 0.41 | 0.13 | 2 | Reed & Allen (1966) |
| 0.15 | 0.03 | 1 | Reed & Jovanovic (1969) |
| 0.36 | 0.12 | 9 | Review: Gladney et al. (1991) |

Note. Values in [ ] are considered uncertain when compared to recommended values, see text.

Data from the Chicago laboratory in Table 7 for BCR-1 (references for Anders, Gros, Hertogen, Krähenbühl, Laul and coworkers) range from 0.046 – 0.093 ppm Br and could illustrate sample heterogeneity (unlikely for standard rocks) and/or analytical variations over time. In section 2, we discussed the potential problem of halogen losses. Bromine might be particularly susceptible to loss and/or incomplete capture during pyrohydrolysis and incomplete retention during assaying, as well as mobilization during sample irradiation, especially from carbonaceous chondrites of type CI and CM (see discussions in Krähenbühl et al. 1973a,b; Ebihara et al. 1982). This is also seen in the RNAA data from Dreibus et al. for CI- and CM chondrites in Table 8 and 9 which are systematically lower than the instrumental NAA data from the Mainz group (Wulf et al. 1995, Palme & Zipfel 2020). This problem seems to apply as well to the CI-data by Reed & Allen (1966) and we do not use thesefor computing CI-chondrite average values (Table 9). In contrast, Lieberman (1966) used NAA and his rock standard values for Br are higher than recommended values (Table 7). Carbonaceous chondrites analyzed by Lieberman & Ehmann (1967) often show higher Br than other authors, which might indicate calibration issues, however, they are not always higher for ordinary chondrites, Table 8. We therefore do not use their bromine value for the CI-chondrite Orgueil (Table 9).

We compare chondrite data for bromine determined by Clay et al. (2017) to other results in Table 8. Their value for the CM2 chondrite Murchison is within the range of other determinations that are on the low side and were excluded for various reasons. Several authors (e.g., Goles et al. 1967) commented on the generally lower alkali and halogen content of Murchison compared to other CM chondrites, which makes it difficult to gauge the reliability of the halogen data from comparing results from only this meteorite. Results for the CM2 chondrite Murray are also difficult to interpret and Murray is known to have low and variable halogen contents (e.g., Goles et al., 1967) which might be due to weathering because some of the Murray samples were only collected long after its fall.

Clay et al. found 0.397 ppm for Br in Allende (CV3) which is the lowest among the literature Br values in Table 8. A few ordinary chondrite falls were analyzed both by Clay et al. (2017) and Kallemeyn & Wasson (1981). Values by Clay et al. (2017) are essentially the same (Forest Vale H4), about 70% higher (Barwell L6) or about 80% lower (Bishunpur LL) than values by Kallemeyn & Wasson (1981); these could indicate natural sample variations. Kallemeyn & Wasson (1981) used radiochemical NAA, but their Br values are not as low as those reported by e.g., Krähenbühl et al. (1973b). The latter noted that their Br data could be artificially low (section 2). Kallemeyn & Wasson (1981) sealed samples individually for irradiation, and losses during irradiation might have been prevented. Further, procedures used for sample transfer for chemical processing must have minimized losses from carbonaceous chondrites that plagued earlier NAA studies.

**Table 8. Bromine in Chondrite Falls (ppm by mass)***

| Br, ppm | 1σ | N | Class, Meteorite, Year of Fall | Reference; Note |
|---|---|---|---|---|
| **CM Chondrite Falls** | | | | |
| 2.2 | | | CM2 Cold Bokkeveld 1838 | Filby & Ball (1965) |
| 4.5 | 0.3 | 2 | CM2 Cold Bokkeveld 1838 | Kallemeyn & Wasson (1981) |
| [1.34] | 0.75 | 2 | CM2 Cold Bokkeveld 1838 | Lieberman & Ehmann (1967) |
| 1.83 | | 1 | CM2 Cold Bokkeveld 1838 | Mittlefehldt (2002) |
| 3.22 | 0.82 | 2 | CM2 Mighei 1889 | Goles & Anders (1962) |
| 4.5 | | 1 | CM2 Mighei 1889 | Kallemeyn & Wasson (1981) |
| [1.18] | | | CM2 Mighei 1889 | Krähenbühl et al. (1973b) |
| 2.32 | | | CM2 Mighei 1889 | Mittlefehldt (2002) |
| 3.5 | 0.1 | 1 | CM2 Mighei 1889 | Reed & Allen (1966) |
| [0.971] | 0.056 | 1 | CM2 Murchison 1969 | Clay et al. (2017) |
| [0.611] | | 1 | CM2 Murchison 1969 | Dreibus et al. (1979) |
| 1.5 | 0.3 | 2 | CM2 Murchison 1969 | Ehmann et al. (1970) |
| [0.5] | 0.06 | 1 | CM2 Murchison 1969 | Grady et al. (1987); uncert. value |
| 3.6 | 1.4 | 2 | CM2 Murchison 1969 | Kallemeyn & Wasson (1981) |
| [0.42] | | | CM2 Murchison 1969 | Krähenbühl et al. (1973b) |
| 2.42 | | | CM2 Murchison 1969 | Wulf et al. (1995) |
| [0.123]** | 0.011 | 2 | CM2 Murray 1950 | Clay et al. (2017) |
| [0.94]** | | | CM2 Murray 1950 | Dreibus et al. (1979) |
| 0.46** | 0.29 | 2 | CM2 Murray 1950 | Goles et al. (1967) |
| 1.33** | 0.08 | 1 | CM2 Murray 1950 | Grady et al. 1987 |
| 0.7** | | 2 | CM2 Murray 1950 | Kallemeyn & Wasson (1981) |
| [0.53]** | | 1 | CM2 Murray 1950 | Krähenbühl et al. (1973b) |
| 3.2 | | 1 | CM2 Nogoya 1879 | Kallemeyn & Wasson (1981) |
| **CO Chondrite Falls** | | | | |
| 1.33 | | 1 | CO3 Felix 1900 | Anders et al. (1976) |
| 1.4 | | 1 | CO3 Felix 1900 | Goles et al. (1967) |
| 1.3 | 0.28 | 2 | CO3 Felix 1900 | Kallemeyn & Wasson (1981) |
| 1.45 | | 2 | CO3 Lance 1872 | Goles et al. (1967) |
| 1.4 | | 2 | CO3 Lance 1872 | Kallemeyn & Wasson (1981) |
| 2.59 | 0.3 | 1 | CO3 Lance 1872 | Reed & Allen (1966) |
| 1.5 | (sic) | 2 | CO3 Ornans 1868 | Kallemeyn & Wasson (1981) |
| [3.02] | 0.33 | 2 | CO3 Ornans 1868 | Lieberman & Ehmann (1967) |
| 4.45 | 0.33 | 2 | CO3 Ornans 1868 | Takahashi et al. (1978) |
| [0.82] | | 1 | CO3 Warrenton 1877 | Dreibus et al. (1979) |
| 1.1 | | 2 | CO3 Warrenton 1877 | Kallemeyn & Wasson (1981) |
| **CV Chondrite Falls** | | | | |
| 1.48 | | 1 | CV3-ox. Allende 1969 | Anders et al. (1975) |
| 1.6 | | 4 | CV3-ox. Allende 1969 | Bischoff et al. (1988) |
| 1.5 | 0.2 | 2 | CV3-ox. Allende 1969 | Buchanan et al. (1997) |
| [0.397] | 0.023 | 1 | CV3-ox. Allende 1969 | Clay et al. (2017) |
| [0.79] | 0.04 | 1 | CV3-ox. Allende 1969 | Dreibus et al. (1979) |
| 1.60 | 0.29 | 1 | CV3-ox. Allende 1969 | Ebihara et al. (1997) |
| 1.5 | 0.4 | 2 | CV3-ox. Allende 1969 | Haas & Haskin (1991) |
| 1.6 | 0.2 | 1 | CV3-ox. Allende 1969 | Izawa et al. (2010) |
| 1.57 | 0.05 | 6 | CV3-ox. Allende 1969 | Kallemeyn & Wasson (1981) |
| 1.58 | 0.10 | 11 | CV3-ox. Allende 1969 | Kallemeyn et al. (1989) |
| 1.56 | 0.02 | 2 | CV3-ox. Allende 1969 | Kato et al. (2000) |
| 1.16 | 0.16 | 6 | CV3-ox. Allende 1969 | Langenauer & Krähenbühl (1993) |
| 1.5 | | 1 | CV3-ox. Allende 1969 | Mittlefehldt et al. (1996) |
| 1.57 | 0.15 | 4 | CV3-ox. Allende 1969 | Mittlefehldt (2002) |
| 1.55 | | 1 | CV3-ox. Allende 1969 | Morgan et al. (1985) |
| 1.53 | | 1 | CV3-ox. Allende 1969 | Müller (1987) |
| 1.50 | 0.06 | 7 | CV3-ox. Allende 1969 | Ozaki & Ebihara (1987) |
| 1.7 | | 2 | CV3-ox. Allende 1969 | Rubin & Wasson (1987) |
| 1.76 | 0.20 | 2 | CV3-ox. Allende 1969 | Wasson et al. (2013) |
| 1.35 | 0.08 | 1 | CV3-ox. Allende 1969 | Wulf et al. (1995) |
| 1.37 | | 1 | CV3-ox Grosnaja 1861 | Anders et al. (1976) |
| 1.94 | | 1 | CV3-red. Mokoia 1908 | Anders et al. (1976) |
| [4.77] | | 1 | CV3-red. Mokoia 1908 | Lieberman & Ehmann (1967) |
| 0.81 | | 1 | CV3-red.Vigarano 1910 | Anders et al. (1976) |
| 0.85 | | 1 | CV3-red.Vigarano 1910 | Dreibus et al. (1979) |
| 0.9 | | 1 | CV3-red.Vigarano 1910 | Kallemeyn & Wasson (1981) |
| **CK Chondrite Falls** | | | | |
| [0.54] | | 1 | CK4 Karoonda 1930 | Dreibus et al. (1979) |
| [1.87] | | 2 | CK4 Karoonda 1930 | Lieberman & Ehmann (1967) |
| 1.5 | 0.2 | 1 | CK4 Karoonda 1930 | Reed & Allen (1966) |
| **Ungrouped Carbonaceous Chondrites** | | | | |
| 2.8 | 0.2 | 1 | C2-ungr. Tagish Lake 2000 | Brown et al. (2000) |
| 3.87 | | 1 | C2-ungr. Tagish Lake 2000 | Dreibus et al. (2004) |
| **H Chondrite Falls** | | | | |
| 0.19 | 0.07 | 2 | H4 Forest Vale 1942 | Clay et al. (2017) |
| 0.20 | | 1 | H4 Forest Vale 1942 | Kallemeyn et al. (1989) |
| 0.36 | | 1 | H4 Kesen 1850 | Dreibus et al. (1979) |
| 0.40 | | 1 | H4 Kesen 1850 | Kallemeyn et al. (1989) |
| < 0.7 | | 1 | H4 Kesen 1850 | Takaoka et al. (1989) |
| 0.16 | 0.03 | 3 | H5 Allegan 1899 | Reed & Allen (1966) |
| 0.02 | | 1 | H5 Allegan 1899 | Morgan et al. (1985) |
| 0.083 | | 1 | H5 Pantar (light) 1938 | Goles et al. (1967) |
| 0.200 | 0.004 | 1 | H5 Pantar (light) 1938 | Reed & Allen (1966) |
| 0.011 | | 1 | H5 Pantar (light) 1938 | Turner (1965) |
| 0.30 | | 1 | H5 Richardton 1918 | Kallemeyn et al. (1989) |
| [0.20] | | 1 | H5 Richardton 1918 | Lieberman & Ehmann (1967) |
| **L Chondrite Falls** | | | | |
| 0.59** | | 1 | L/LL4 Bjurböle 1899 | Dreibus et al. (1979) |
| 0.50** | | 1 | L/LL4 Bjurböle 1899 | Kallemeyn et al. (1989) |
| 0.16 | | 1 | L/LL4 Bjurböle 1899 | Turner (1965) |
| 2.60 | 0.57 | 2 | L3.6 Khohar 1910 | Kallemeyn et al. (1989) |
| 5.57 | 2.12 | 2 | L3.6 Khohar 1910 | Keays et al. (1971) |
| 2.04 | 0.14 | 2 | L4 Fukotomi 1882 | Keays et al. (1971) |
| 4.14 | 0.80 | 4 | L4 Fukotomi 1882 | Takaoka et al. (1989) |
| 0.47 | | 1 | L4 Saratov 1918 | Hey (1966) |
| 0.60 | | 1 | L4 Saratov 1918 | Kallemeyn et al. (1989) |
| 0.47 | | 1 | L4 Saratov 1918 | Selivanov (1940) |
| 1.10 | | 1 | L4Tennasilim 1872 | Kallemeyn et al. (1989) |
| [1.05] | 0.08 | 2 | L4Tennasilim 1872 | Keays et al. (1971) |
| 0.18 | | 1 | L6 Alfianello 1883 | Dreibus et al. (1979) |
| 0.30 | | 1 | L6 Alfianello 1883 | Kallemeyn et al. (1989) |
| 0.42 | 0.08 | 1 | L6 Barwell 1965 | Clay et al. (2017) |
| 0.25 | | 1 | L6 Barwell 1965 | Kallemeyn et al. (1989) |
| 0.20** | | 1 | L6 Bruderheim 1960 | Dreibus et al. (1979) |
| 0.15** | 0.03 | 6 | L6 Bruderheim 1960 | Goles et al. (1967) |
| 1.35** | 0.16 | 1 | L6 Bruderheim 1960 | Ebihara et al. (1997) |
| 0.84** | 0.13 | 1 | L6 Bruderheim 1960 | Kato et al. (2000) |
| 0.17** | 0.08 | 5 | L6 Bruderheim 1960 | Reed & Allen (1966) |
| 0.106** | 0.02 | 6 | L6 Bruderheim 1960 | Keays et al. (1971) |
| 0.028** | 0.003 | 2 | L6 Bruderheim 1960 | Lieberman & Ehmann (1966) |
| 0.20** | | 1 | L6 Bruderheim 1960 | Merrihue (1966) |
| 1.27** | 0.42 | 2 | L6 Bruderheim 1960 | Wyttenbach et al. (1965) |
| 0.10 | 0.04 | 2 | L6 Harleton 1961 | Reed & Allen (1966) |
| [0.29] | | 1 | L6 Harleton 1961 | Lieberman & Ehmann (1967) |
| 1.00** | | 1 | L6 Holbrook 1912 | Filby & Ball (1965) |
| 0.72** | 0.16 | 2 | L6 Holbrook 1912 | Goles et al. (1967) |
| 0.99** | 0.06 | 5 | L6 Holbrook 1912 | Langenauer & Krähenbühl (1993) |
| 1.50** | 0.55 | 3 | L6 Holbrook 1912 | Reed & Allen (1966) |
| 0.32 | | 1 | L6 Kunashak 1949 | Lieberman & Ehmann (1967) |
| 0.22 | | 1 | L6 Kunashak 1949 | Wyttenbach et al. (1965) |
| 0.19 | | 1 | L6 Leedey 1943 | Dreibus et al. (1979) |
| 0.10 | | 1 | L6 Leedey 1943 | Kallemeyn et al. (1989) |
| 0.27 | | 1 | L6 Mocs 1860 | Dreibus et al. (1979) |
| 0.03 | | 1 | L6 Mocs 1860 | Goles et al. (1967) |
| [11.4] | | 1 | L6 Mocs 1860 | Behne (1953) |
| 0.15 | 0.09 | 3 | L6 Mocs 1860 | Wyttenbach (1967) |
| 0.05 | 0.02 | 2 | L6 Mocs 1860 | Wyttenbach et al. (1965) |
| 0.02 | | 1 | L6 Peace River 1963 | Goles et al. (1967) |
| 0.16 | | 2 | L6 Peace River 1963 | Lieberman & Ehmann (1967) |
| **LL Chondrite Falls** | | | | |
| [0.12] | | 1 | LL3 Bishunpur 1895 | Clay et al. (2017) |
| 0.70 | | 1 | LL3 Bishunpur 1895 | Kallemeyn et al. (1989) |
| 0.80 | | 1 | LL3 Ngawi 1883 | Kallemeyn et al. (1989) |
| 0.60 | | 1 | LL3 Semarkona 1940 | Kallemeyn et al. (1989) |
| 0.87 | | | LL3.4 Chainpur 1907 | Goles et al. (1967) |
| 1.0 | | | LL3.2 Krymka 1946 | Kallemeyn et al. (1989) |
| 1.98 | 0.24 | 2 | LL3.2 Krymka 1946 | Keays et al. (1971) |
| 2.26 | | 1 | LL3.6 Parnalee 1857 | Wyttenbach (1967) |
| 1.0 | | 2 | LL4 Hamlet 1959 | Kallemeyn et al. (1989) |
| 1.85 | 0.4 | 2 | LL4 Hamlet 1959 | Keays et al. (1971) |
| [0.025] | 0.009 | 3 | LL5 Chelyabinsk 2013 | Clay et al. (2017) |
| 0.20 | | 1 | LL6 Dhurmsala 1860 | Kallemeyn et al. (1989) |
| 0.66 | | 1 | LL6 Dhurmsala 1860 | Wyttenbach et al. (1965) |
| **K Chondrite Falls** | | | | |
| [0.013] | 0.001 | 2 | K3 Kakangari 1890 | Clay et al. (2017) |
| 0.90 | | 1 | K3 Kakangari 1890 | Weisberg et al. (1996) |
| **EH Chondrite Falls** | | | | |
| 3.0 | | 1 | EH3 Qingzhen 1979 | Grossman et al. 1985 |
| 2.56 | | 1 | EH4 Abee 1952 | Dreibus et al. (1979) |
| 2.07 | | 1 | EH4 Abee 1952 | Filby & Ball (1965) |
| 3.34 | 0.03 | 1 | EH4 Abee 1952 | Goles et al. (1967) |
| 3.63 | | 1 | EH4 Abee 1952 | Hertogen et al. (1983) |
| 1.75 | 0.22 | 2 | EH4 Abee 1952 | Lieberman & Ehmann (1967) |
| 3.60 | | 1 | EH4 Abee 1952 | Merrihue (1966) |
| 3.50 | 0.05 | 1 | EH4 Abee 1952 | Reed & Allen (1966) |
| [0.327] | 0.03 | 3 | EH4 Indarch 1891 | Clay et al. (2017) |
| 4.80 | | 1 | EH4 Indarch 1891 | Goles et al. (1967) |
| 2.40 | 0.14 | 2 | EH4 Indarch 1891 | Kallemeyn & Wasson (1986) |
| 6.50 | 0.30 | 1 | EH4 Indarch 1891 | Reed & Allen (1966) |
| [0.14] | 0.02 | 1 | EH5 St. Marks 1903 | Clay et al. (2017) |
| 0.6 – 1.3 | | | EH5 St. Marks 1903 | Kallemeyn & Wasson (1986) |
| **EL Chondrite Falls** | | | | |
| [0.020] | 0.002 | 2 | EL6 Daniels's Kuil 1868 | Clay et al. (2017) |
| 0.54 | | 1 | EL6 Daniels's Kuil 1868 | Hertogen et al. (1983) |
| 0.95 | | 1 | EL6 Hvittis 1901 | Goles et al. (1967) |
| 0.88 | | 1 | EL6 Hvittis 1901 | Hertogen et al. (1983) |
| 0.80 | | 2 | EL6 Hvittis 1901 | Kallemeyn & Wasson (1986) |
| 1.50 | 0.10 | 1 | EL6 Hvittis 1901 | Reed & Allen (1966) |
| 1.04 | | 1 | EL6 Hvittis 1901 | Wyttenbach et al. (1965) |
| 1.6 | 0.11 | 1 | EL6 Neuschwanstein 2002 | Zipfel et al. (2010) |

*Only meteorite falls with more than one analyses for Cl are listed.

** Notes for various meteorites: Bjurböle fell into frozen lake. Bruderheim is known to be heterogeneous. Holbrook samples were collected for 50+ years after fall, weathering is noted on some specimens. Murray is known to have lower volatiles than other CM; some samples were only collected long after its after fall.

The Br determinations in ordinary and enstatite chondrites are difficult to interpret. There are no obvious indications for systematic variations of results by different groups. Sample heterogeneity certainly is an issue for ordinary chondrites where halogens were scavenged into secondary phosphates (e.g., review by Brearley & Jones 2018).

Table 9 summarizes CI-chondrite bromine concentrations. Considering Orgueil first, we use values from seven studies mainly from RNAA by Ebihara et al. (1982), Goles et al. (1967), Grossman et al. (1985), Kallemeyn & Wasson (1981), Mittlefehldt (2002), Palme & Zipfel (2021), and Takahashi et al. (1978). We did not include the very recent data by Ebihara & Sekimoto (2019), but their results agree very well. We also included data by Burnett et al. (1989) from proton-induced X-ray emission (PIXE) spectroscopy which is sensitive to the outmost 30-microns of sample surfaces. Giving the same statistical weight to the nine studies, we obtain an average Br concentration of 3.34±0.36 ppm for Orgueil.

**Table 9 Bromine in CI-Chondrites**

| Br, ppm | 1σ | N | Reference |
|---|---|---|---|
| | | | **Alais 1806** |
| 2.26 | | 1 | Ebihara et al. (1982) |
| 2.4 | | 2 | Kallemeyn & Wasson (1981) |
| [0.93] | | 1 | Krähenbühl et al. (1973b) |
| 4.39 | 0.82 | 2 | Palme & Zipfel (2021) |
| **3.02** | **1.19** | **3** | **Average Alais** |
| | | | **Ivuna 1939** |
| 4.95 | 0.96 | 3 | Burnett et al. (1989) |
| [3.91] | 0.16 | 1 | Dreibus et al. (1979) |
| 5.05 | | 1 | Ebihara et al. (1982) |
| 5.10 | | 1 | Goles et al. (1967) |
| [2.35] | 0.18 | 3 | Krähenbühl et al. (1973b) |
| 5.68 | 1.13 | 3 | Palme & Zipfel (2021) |
| [0.5; 11] | | | Reed & Allen (1966) |
| **4.96** | **0.60** | **5** | **Average Ivuna** |
| | | | **Orgueil 1864** |
| 3.10 | 0.32 | 3 | Burnett et al. (1989) |
| [0.189] | 0.071 | 2 | Clay et al. (2017) |
| [2.53] | 0.2 | 4 | Dreibus et al. (1979) |
| 3.58 | 0.59 | 3 | Ebihara et al. (1982) |
| 3.58 | 0.29 | 3 | Goles et al. (1967) |
| 3.30 | | 1 | Grossman et al. (1985) |
| 3.7 | 0.1 | 3 | Kallemeyn & Wasson (1981) |
| [1.45] | 0.11 | 3 | Krähenbühl et al. (1973b) |
| [5.72] | | 1 | Lieberman & Ehmann (1967) |
| 3.08 | | 1 | Mittlefehldt et al. (2002) |
| 2.68 | 0.54 | 8 | Palme & Zipfel (2021) |
| [2.0] | | 1 | Reed & Allen (1966) |
| 3.67 | | 1 | Takahashi et al. 1978 |
| **3.34** | **0.34** | **9** | **Average Orgueil** |
| | | | **Tonk 1911** |
| 9.0 | 0.9 | 1 | Palme & Zipfel (2021) |

Note: Values in [ ] excluded from averages, see text.

Bromine concentrations in Alais from NAA show larger variations among different studies. There is also larger variance within individual studies and in part these variations may reflect not only analytical difficulties but also sample heterogeneity. It has been known for some time that bromine concentrations in Alais are lower and in Ivuna are higher than in Orgueil. The single analyses for the Tonk chondrite with high bromine would indicate that Tonk and Ivuna are similar in this regard. Reed & Allen (1966) found 11 ppm Br in one Ivuna sample which is suspiciously high and most easily explained by contamination, but instances of halogen-rich samples might just reflect the upper end of the true distribution for sampling.

Kallemeyn & Wasson (1981) exclude their bromine value for Alais in the computation of the CI-chondrite mean, but we take the weighted average from Alais, Ivuna and Orgueil and obtain of 3.77±0.90 ppm bromine, where the uncertainty is the standard deviation from the mean. This uncertainty well reflects the range among the three CI-chondrites and the spread seen among individual analyses. The concentration corresponds to (12.3±2.9) Br per $10^6$ silicon atoms. Anders & Ebihara (1982) and Anders & Grevesse (1989) suggest 3.56 ppm, derived from Br/In and Br/Cd abundance ratios in E4 and CV3 chondrites which presumably contain unfractionated volatile elements. We refrain from using the element ratios here because the In and Cd abundances are also plagued with uncertainties.

### 3.2.3. Iodine

Iodine analyses are very sparse and the few data for rock standards, chondrites, and CI-chondrites are in Tables 10-12. A comparison of rock standards (Table 10) is difficult because several groups analyzing meteorites did not analyze or report these standards (e.g., Clay et al. 2017, Goles et al. 1967). Dreibus et al. (1979) and Langenauer et al. (1992) used the same analytical techniques and the only value in common (for BCR-1) agrees well, but both are well below the recommended rock standard value of 180 ppb from Gladney et al. (1990). However, there are very few standard measurements for BCR-1 listed by Gladney et al. Values for other rock standards show no trends for systematically lower values; e.g., Dreibus et al. report higher values for the G-1 and W-1 standards than Gladney et al. (1991) recommended.

**Table 10. Iodine in Rock Standards (ppm by mass) also Analyzed in Meteorite Studies, and Literature Survey**

| I, ppm | 1σ | N | Reference |
|---|---|---|---|
| **AGV-1 Andesite Rock Standard** | | | |
| 0.24 | 0.04 | 2 | Langenauer et al. (1992) |
| 0.266 | 0.030 | 5 | Review: Gladney et al.(1992) |
| **BCR-1 Columbia River Basalt Rock Standard** | | | |
| 0.05 | 0.01 | 1 | Dreibus et al. (1977,1979) |
| 0.041 | 0.008 | 2 | Langenauer et al. (1992) |
| 0.18 | 0.008 | | Review: Gladney et al. (1990) |
| **G-1 Granite Rock Standard** | | | |
| 0.07 | | 1 | Dreibus et al. (1979) |
| 0.035 | | 1 | Review: Gladney et al. (1991) |
| **W-1 Granite Rock Standard** | | | |
| 0.07 | | 1 | Dreibus et al. (1979) |
| 0.054 | | 2 | Review: Gladney et al. (1991) |

Iodine is prone to redistribution during aqueous alterations and metamorphism in meteorites. In ordinary chondrite iodine is associated with olivine, pyroxenes, metal, sulfides, and secondary phases such as apatite, sodalite, nepheline (e.g., Goswami et al. 1998; Brazzle et al. 1999; Lewis & Jones 2016; Ward et al. 2017), and in enstatite chondrites iodine is apparently sequestered in enstatite crystals or as fine inclusions within enstatite (Kehm et al. 1994). Iodine in CI-chondrites was measured by Clay et al. (2017), Dreibus et al. (1979), Goles et al. (1967), and Reed & Allen (1966) (Table 12). Their results for other chondrites are given in Table 11 together with other studies. As described above, several iodine concentrations were obtained in studies for I-Xe dating from $^{128}Xe$ derived from irradiated samples, and more often (but not always) the iodine contents from these studies (e.g., Crabb et al. 1982, Hohenberg et al. 1981, Jordan et al. 1980, in Table 11) are lower than those measured by other means. Data by Clay et al. (2017) from $^{128}Xe$ measurements also appear systematically low for several meteorites (Table 11).

**Table 11. Iodine in Chondrites (Falls Only)**

| I, ppm | 1σ | N | Class, Meteorite, Year of Fall | Reference; Note |
|---|---|---|---|---|
| **CM Chondrite Falls** | | | | |
| 0.31 | 0.06 | 1 | CM2 Mighei 1889 | Goles & Anders (1962) |
| 0.48 | | 1 | CM2 Mighei 1889 | Goles et al. (1967) |
| 0.55 | 0.04 | 1 | CM2 Mighei 1889 | Reed & Allen (1966) |
| [0.065] | 0.003 | 1 | CM2 Murchison 1969 | Clay et al. (2017) |
| 0.19 | | | CM2 Murchison 1969 | Dreibus et al. (1979) |
| [0.014]** | 0.002 | 2 | CM2 Murray 1950 | Clay et al. (2017) |
| 0.66** | | 1 | CM2 Murray 1950 | Dreibus et al. (1979) |
| 0.23** | 0.08 | 3 | CM2 Murray 1950 | Goles & Anders 1962 |
| 0.3** | | 1 | CM2 Murray 1950 | Goles et al. (1967) |
| **CO Chondrite Falls** | | | | |
| [0.058] | | 1 | CO3 Lance 1872 | Crabb et al. (1982) |
| 0.17 | | 1 | CO3 Lance 1872 | Goles et al. (1967) |
| 0.11 | 0.01 | 11 | CO3 Lance 1872 | Reed & Allen (1966) |
| [0.062] | | 1 | CO3 Warrenton 1877 | Crabb et al. (1982) |
| 0.11 | | 1 | CO3 Warrenton 1877 | Dreibus et al. (1979) |
| **CV Chondrite Falls** | | | | |
| [0.035] | 0.001 | 3 | CV3-ox. Allende 1969 | Clay et al. (2017) |
| 0.158 | | | CV3-ox. Allende 1969 | Crabb et al. (1982) |
| 0.16 | 0.02 | | CV3-ox. Allende 1969 | Dreibus et al. (1979) |
| 0.16 | 0.02 | | CV3-ox. Allende 1969 | Ebihara et al. (1986) |
| 0.212 | 0.03 | | CV3-ox. Allende 1969 | Ebihara et al. (1997) |
| 0.120 | 0.003 | | CV3-ox. Allende 1969 | Heumann & Weiss (1986) |
| 0.205 | 0.05 | 2 | CV3-ox. Allende 1969 | Kato et al. (2000) |
| 0.19 | 0.03 | 7 | CV3-ox. Allende 1969 | Langenauer & Krähenbühl (1993) |
| 0.166 | 0.018 | 2 | CV3-ox. Allende 1969 | Ozaki & Ebihara (2007) |
| [>0.034] | | 1 | CV3-red. Vigarano 1910 | Crabb et al. (1982) |
| 0.3 | | 1 | CV3-red. Vigarano 1910 | Dreibus et al. (1979) |
| **CK Chondrite Falls** | | | | |
| [0.019] | | 1 | Karoonda 1930 | Crabb et al. (1982) |
| 0.52 | | 1 | Karoonda 1930 | Dreibus et al. (1979) |
| **H Chondrite Falls** | | | | |
| 0.022 | 0.002 | 2 | H4 Forest Vale 1942 | Clay et al. (2017) |
| 0.06 | | | H4 Kesen 1850 | Dreibus et al. (1979) |
| 0.069 | 0.004 | 1 | H4 Allegan 1899 | Reed & Allen (1966) |
| 0.063 | 0.004 | 1 | H4 Beardsley 1929 | Goles & Anders (1962) |
| 0.024 | 0.011 | 2 | H4 Gao-Guenie 1960 | Heumann & Weiss (1986) |
| 0.028 | 0.006 | 3 | H5 Richardton 1918 | Goles & Anders (1962) |
| 0.023 | | 1 | H5 Richardton 1918 | Merrihue (1966) |
| **L Chondrite Falls** | | | | |
| >0.018 | | 1 | L/LL4 Bjurböle | Brazzle et al. (1999) |
| 0.07** | | 1 | L/LL4 Bjurböle | Dreibus et al. (1979) |
| >0.032** | 0.0032 | 1 | L/LL4 Bjurböle | Drozd & Podosek (1976) |
| >0.016** | | 1 | L/LL4 Bjurböle | Hohenberg et al. (1981) |
| >0.024** | | 1 | L/LL4 Bjurböle | Jordan et al. (1980) |
| 0.016** | 0.003 | 1 | L/LL4 Bjurböle | Turner (1965) |
| 0.023** | 0.006 | 2 | L6 Bruderheim 1960 | Clark et al. (1967) |
| 0.04** | | 1 | L6 Bruderheim 1960 | Dreibus et al. (1979) |
| 0.005; 0.022** | | 2 | L6 Bruderheim 1960 | Goles & Anders (1962) |
| 0.007** | 0.001 | 2 | L6 Bruderheim 1960 | Goles et al. (1967) |
| 0.024** | 0.014 | 1 | L6 Bruderheim 1960 | Ebihara et al. (1997) |
| 0.032** | 0.012 | 1 | L6 Bruderheim 1960 | Kato et al. (2000) |
| 0.0078** | | 1 | L6 Bruderheim 1960 | Merrihue (1966) |
| 0.074; <0.45** | | 2 | L6 Bruderheim 1960 | Reed & Allen (1966) |
| 0.016** | 0.003 | 3 | L6 Bruderheim 1960 | Turner (1963) |
| 0.104** | 0.005 | 5 | L6 Holbrook 1912 | Langenauer & Krähenbühl (1993) |
| 0.03** | 0.002 | 1 | L6 Holbrook 1912 | Reed & Allen (1966) |
| 0.08 | | 1 | L6 Mocs 1860 | Dreibus et al. (1979) |
| 0.05 | 0.03 | 1 | L6 Mocs 1860 | Goles & Anders (1962) |
| 0.014 | 0.004 | 2 | L6 Mocs 1860 | Goles et al. (1967) |
| 0.058 | 0.005 | 1 | L6 Modoc 1905 | Ebihara et al. (1986) |
| 0.018 | 0.001 | 2 | L6 Peace River 1963 | Clark et al. (1967) |
| 0.013 | | 1 | L6 Peace River 1963 | Goles et al. (1967) |
| **LL Chondrite Falls** | | | | |
| 0.009 | 0.002 | 2 | LL3 Bishunpur 1895 | Clay et al. (2017) |
| 0.20 | | 1 | LL3.4 Chainpur | Goles et al. (1967) |
| 0.048 | 0.004 | 3 | LL5 Chelyabinsk 2013 | Clay et al. (2017) |
| 0.13 | 0.03 | 1 | LL6 Saint Severin 1966 | Ebihara et al. (1986) |
| **EH Chondrite Falls** | | | | |
| 0.45 | | 1 | EH4 Abee 1952 | Dreibus et al. (1979) |
| 0.145 | 0.007 | 2 | EH4 Abee 1952 | Goles & Anders (1962) |
| 0.3 | 0.08 | 1 | EH4 Abee 1952 | Merrihue (1966) |
| [<0.18] | | 1 | EH4 Abee 1952 | Reed & Allen (1966) |
| [0.062] | 0.006 | 3 | EH4 Indarch 1891 | Clay et al. (2017) |
| 0.21 | 0.05 | 3 | EH4 Indarch 1891 | Busfield et al. (2008) |
| 0.45 | | 1 | EH4 Indarch 1891 | Dreibus et al. (1979) |
| 0.27 | 0.05 | 2 | EH4 Indarch 1891 | Goles & Anders (1962) |
| 0.31 | | 1 | EH4 Indarch 1891 | Goles et al. (1967) |
| [<0.18] | | 1 | EH4 Indarch 1891 | Reed & Allen (1966) |
| [0.011] | 0.002 | 1 | EH5 St. Marks 1903 | Clay et al. (2017) |
| 0.082 | 0.026 | 2 | EH5 St. Marks 1903 | Goles & Anders (1962) |
| **EL Chondrite Falls** | | | | |
| 0.0016 | 0.0002 | 2 | EL6 Daniel's Kuil | Clay et al. (2017) |
| 0.017-0.089 | | 2 | EL6 Hvittis 1901 | Goles et al. (1967) |
| 0.074 | 0.017 | 1 | EL6 Hvittis 1901 | Reed & Allen (1966) |
| 0.061 | 0.013 | | EL6 Khaipur 1873 | Busfield et al. (2008) |
| 0.14 | 0.014 | 1 | EL6 Neuschwanstein 2002 | Zipfel et al. (2010) |

Note: Values in [ ] excluded from averages, see text. ** Bjurböle fell into frozen lake. Bruderheim is known to be heterogeneous. Holbrook samples were collected for over 50 years after fall, weathering is noted on some specimens. Murray has lower volatiles than other CM; some samples were only collected long time after fall.

Table 12 summarizes the iodine contents for two CI-chondrites. For Ivuna, Goles et al. (1967) found a significantly large iodine content of 11 ppm, which they ascribe to possible contamination or sample heterogeneity. Data from Reed & Allen (1966) are lower than those of Dreibus et al. (1979), but comparison of the Reed & Allen data for other meteorites (Table 11) does not reveal systematically lower values.

Given the large differences in results, we computed the geometric mean for each meteorite and the straight mean as done for the other elements (where arithmetic and geometric average are essentially identical). This is a better approach to avoid skewing the average toward a higher value. The weighted average (as calculated for the other elements) for the CI-chondrites Ivuna (and Orgueil yield 0.77±0.31 ppm for iodine. We adopt an uncertainty of 40% which reflects the large concentration differences between the two meteorites, and the individual studies of each meteorite. Without further measurements on well defined, representative samples the iodine abundance from CI-chondrites remains poorly known. Ebihara & Sekimoto (2019) found iodine abundances that are at the higher end and comparable to the values by Dreibus et al. (1979), however, several other meteorites analyzed by Ebihara & Sekimoto show anomalously large iodine concentration and reasons for this need to be understood first. However, it is unlikely that the iodine concentration would be up to a factor of ten less as suggested from the measurement by Clay et al. (2017) for Orgueil or that it is a factor of ten higher as suggested by the measurement by Goles et al. (1967) for Ivuna. While the iodine value is uncertain within a factor of two, we discuss other evidence below that an iodine content of 0.77 ppm is a plausible estimate for CI-chondrites.

**Table 12. Iodine in CI-Chondrites**

| I ppm | 1σ | N | | |
|---|---|---|---|---|
| | | **Ivuna 1939** | | |
| 1.73 | 0.2 | 1 | | Dreibus et al. (1979) |
| [11.2] | | 1 | | Goles et al. (1967) |
| 0.92 | 0.37 | 3 | | Reed & Allen (1966) |
| 1.33 | 0.57 | 2 | | Arithmetic Average Ivuna |
| 1.26 | +0.71 -0.46 | 2 | | Geometric Average Ivuna |
| | | | | |
| | | **Orgueil 1864** | | |
| [0.057] | 0.007 | 2 | | Clay et al. (2017) |
| 0.56 | 0.12 | 3 | | Dreibus et al. (1979) |
| 0.40 | | 1 | | Goles et al. (1967) |
| 0.23 | | 2 | | Reed & Allen (1966) |
| 0.40 | 0.17 | 3 | | Arithmetic Average Orgueil |
| 0.37 | +0.15 -0.10 | 3 | | Geometric Average Orgueil |

Note. Values in [ ] excluded from averages, see text.

## 3.3. Halogen Abundances in CI-Chondrites and comparison to other carbonaceous chondrites and nuclear systematics

Table 13 gives our recommended average CI-chondrite concentrations and abundances relative to N(X) = $10^6$ Si atoms and on the logarithmic atomic scale A(X) normalized to log N(H) =12. The linear scale relative to silicon is related to the log scales as A(X/) = log N(X) + 1.51.

**Table 13. Weighted Average Halogen Concentrations and Abundances in CI-Chondrites**

| | Concentration ppm by mass (±1σ) | Number rel. Si=$10^6$ atoms | A(X) = 12+log(X/H) (meteoritic) |
|---|---|---|---|
| F | 92±20 | 1270±270 | 4.61±0.09 |
| Cl | 717±110 | 5290±810 | 5.23±0.06 |
| Br | 3.77±0.90 | 12.3±2.9 | 2.60±0.09 |
| I | 0.77±0.31 | 1.59±0.64 | 1.71±0.15 |

Note: Average of the mean abundances of the individual CI-Chondrites (Tables 3, 6, 9, 12) were weighted by number of studies per CI-Chondrite to obtain the CI-chondrite group abundances. For iodine, see text.

We compare halogen abundances of CI-chondrites to other studies and to other carbonaceous chondrites in Figure 2. Overall, the CI-chondrites have the highest halogen/silicon ratios among the carbonaceous chondrites and it has been known for a long time that volatile elements in CI-chondrites have higher abundances (relative to silicon or some other refractory element) than in other carbonaceous chondrite groups (e.g., Wasson & Kallemeyn 1988, Palme et al. 2014).

Our fluorine values for CI- and CM-chondrites are higher than previous values which is mainly due to the incorporation of new measurements and not discarding the values by Greenland & Lovering because no good reason exists to do so. Except for the values by Clay et al. (2017), chlorine values are comparable to the data by Wasson & Kallemeyn (1988) and Dreibus et al. (1979). Several measurements compare well to data by Kallemeyn & Wasson (1988) but the values by Dreibus et al. (1979) are systematically lower (which is why we did not include their Br data in our evaluation as described above). As for Cl, the Br data for CI-chondrites by Clay et al. (2017) are far too low, their average from the CM-chondrites Murray and Murchison is more consistent with other results, but their value for CV-chondrites is also quite low.

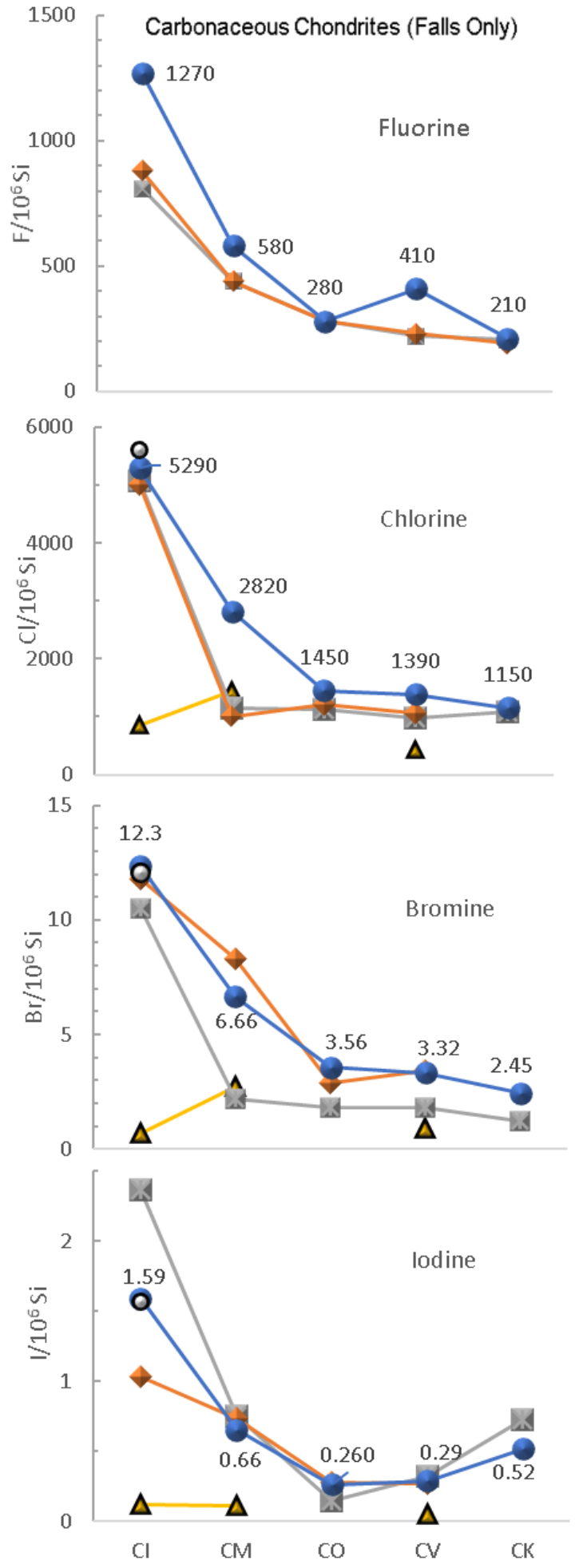

**Figure 2**. Average atomic abundances of F, Cl, Br, and I in different carbonaceous chondrite groups relative to $10^6$ Silicon. Only data for meteorite falls were used. Large blue circles with numerical values: This work. Triangles: Clay et al. (2017). Grey squares: Dreibus et al. (1979). Open circle: Ebihara & Sekimoto (2019). Diamonds: Wasson & Kallemeyn (1988). For uncertainties, see Table 13.

Higher abundance ratios in CM-chondrites than in CI-chondrites by Clay et al. (2017) seem suspicious because other volatile elements are generally lower in CM chondrites (lower element/Si ratios) than in CI-chondrites indicating that there were problems in the study by Clay et al. (2017) as discussed above. The iodine abundances for carbonaceous chondrite groups by Clay et al. (2017) are up to ten time lower than reported by other groups. Our evaluated iodine abundance of I/$10^6$ Si = 1.58±0.64 is between the values from Dreibus et al. (1979) and Wasson &

Kallemeyn (1988) and coincides with that of Ebihara & Sekimoto (2019). We did not plot the uncertainties in Figure 2 to avoid cluttering the graphs. But especially for iodine the CI-chondrite abundances are within the range of the values by Kallemeyn & Wasson (1988) and those of Dreibus et al. (1979), but the datum by Clay et al. (2017) is far below even within uncertainties (Figure 2). The smaller spread in iodine abundances for the other carbonaceous chondrite groups mainly reflects the small number of analyses, and the difficulties associated with the analyses of CI-chondrites described above.

One traditional test for elemental abundances is how they fit into the isotopic abundance distribution of the odd-numbered nuclides (see Figure 3, Suess 1947, Burnett et al. 1989). In order to plot the halogens, their terrestrial isotopic compositions were assumed; meteoritic isotopic compositions vary from terrestrial values at per-mil levels which does not affect the conclusions drawn from this plot. The halogen abundances fit well into the abundance distribution (although not too much can be gauged for F because the neighboring elemental and isotopic O and Ne abundances are relatively uncertain and still debated). Also shown in Figure 3 are the Cl, Br, and I values from Clay et al. (2017) which plot far below the smooth abundance distribution curve, indicating that their proposed halogen abundances are not representative for solar system abundances. Overall, much lower concentrations than currently accepted values for the halogens are not consistent with the general elemental and nuclide abundance systematics.

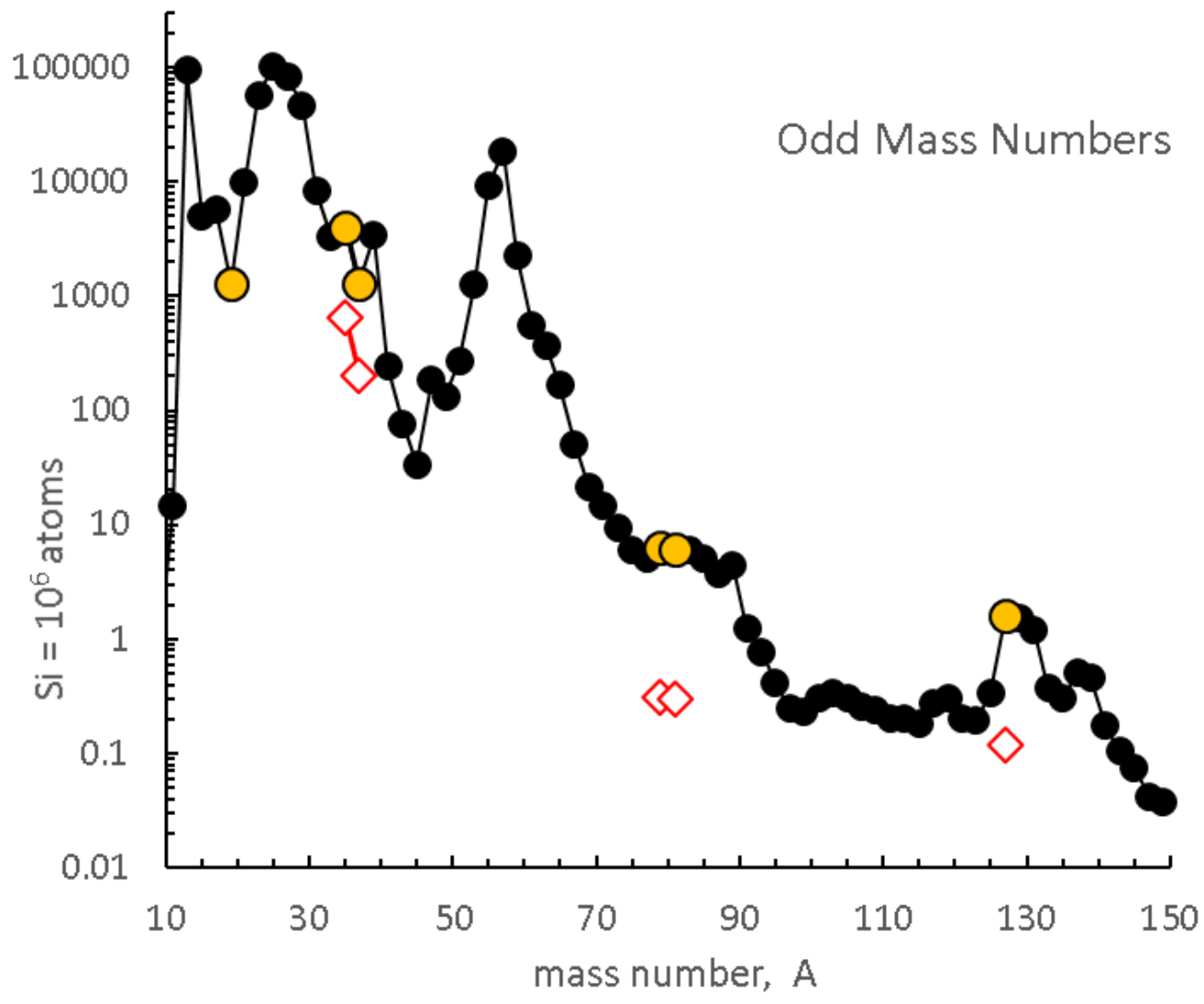

**Figure 3**: Abundances of the odd-numbered isotopes vs. mass number (normalized to Si = $10^6$ atoms). Recommended halogen abundances are shown as lighter circles (orange in the color version). Values for Cl, Br, and I proposed by Clay et al. (2017; red open diamonds) are far below the smooth abundance curve.

## 4. Halogens in the Sun, Other Astronomical Environs, and Recommended Abundances

In this section we summarize some results for the halogens in the sun, asteroid Ryugu, comet 67P/Churyumov-Gerasimenko, other stars, the interstellar medium (ISM), HII regions, planetary nebulae (PNe), and supernova remnants (SNRs). Many of the environments outside the solar system have relative elemental abundances close to solar system values if chemical fractionations and nucleosynthetic production mechanisms are considered, and generally our derived abundances for CI-chondrites compare well.

Fluorine and chlorine abundances in the sun are determined from sunspot data, for F: Hall & Noyes (1969). Maiorca et al. (2014); for Cl: Hall and Noyes (1972), Maas et al. (2016). These values are typically adopted in many solar abundance compilations (see Lodders (2020) for a comprehensive data summary on historical solar elemental abundance estimates).

## 4.1. Fluorine

Table 14 and Figure 4 summarize fluorine abundances, which are given as A(F) = log ϵ(F) = 12+ log (F/H). Fluorine abundances in the sun and stars in the solar neighborhood range from $4.2 \leq A(F) \leq 4.8$, and the meteoritic value is A(F) = 4.61 ± 0.09. The CI-chondritic F/$10^6$ Si ratio was scaled to the H-based abundance scale by using the Si/H ratio of the photosphere. Note that previously the meteoritic value used to be smaller than the solar sunspot value while this situation is now reversed. The earlier F abundance from sun-spot spectra was about A(F) = 4.56 on the scale for log A(H) = 12 (Hall & Noyes 1969). Maiorca et al. (2014) re-analyzed a 1982 sunspot spectrum for HF using updated molecular parameters and found a fluorine abundance of A(F) = 4.40±0.25, which Abia et al. (2015) confirmed. Another solar-system estimate for the fluorine abundance is from the HF/oxygen ratios observed in comet 67P/Churyumov-Gerasimenko, where O was derived from $H_2O+2CO_2+CO+2\ O_2$ (Dhooghe et al. 2017, Rubin et al. 2019). The uncertainties are large, but the nominal cometary F abundance is more consistent with CI-chondrites than sunspots. Selective volatilization of more volatile oxygen from ice would decrease the F/O ratio and F abundance observed in the comet.

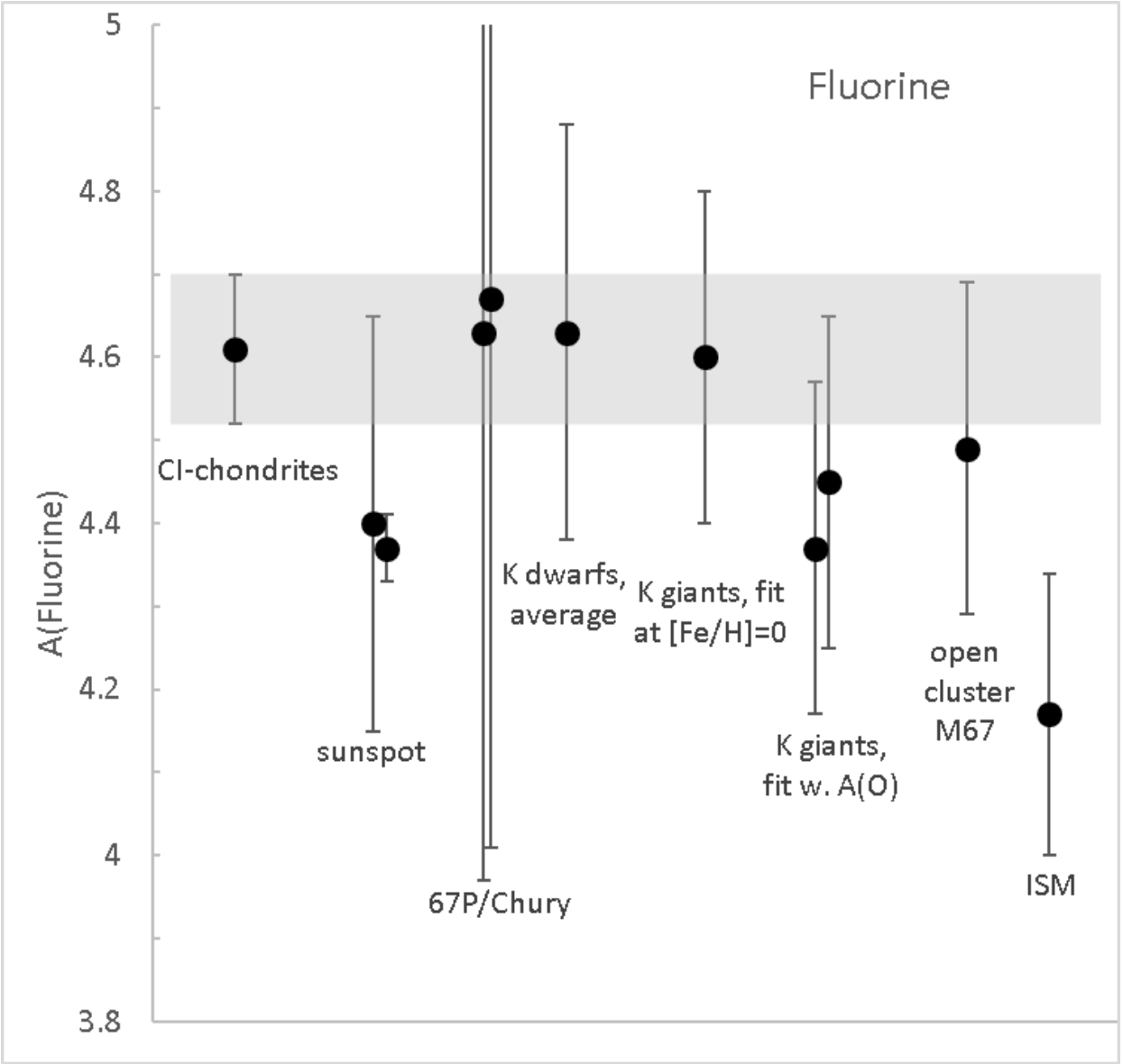

**Figure 4**. Fluorine abundances in CI-chondrites, sunspots and other objects as indicated. The shaded band is the range for CI-chondrites. For data sources see Table 14. If no uncertainties were given, an uncertainty of 0.2 dex was assumed.

Maiorca et al. (2014) also derived A(F)=4.49±0.20 for the solar-metallicity open cluster M67. Recio-Blanco et al (2012) determined fluorine abundances for nine main-sequence dwarf stars in the solar neighborhood. For six stars (excluding the BY Dra and RS CVn variables among them) with slightly sub-solar to solar metallicity (-0.37 < [Fe/H] < -0.04) the average is A(F) = 4.60 ±0.13, however, these results are superseded by analyses using updated excitation potentials for the HF feature.

The eight dwarf stars with -012< [Fe/H] <0.26 in the sample by Pilachowski & Pace (2015) using the new excitation potentials average to A(F) = 4.73 ±0.30; omitting LHS1453 with A(F) = 5.1 gives an average A(F)=4.63 ±0.25. Two of their stars within 0.06 dex of solar metallicity have A(F) = 4.68; so all values are higher than the sunspot value and closer to the meteoritic value. Pilachowski & Pace (2015) fitted their sample of solar neighborhood thin disk giants to [F/ Fe] = –0.105[Fe/H] + 0.20, suggesting "an average fluorine abundance of [F/Fe] = +0.20 (log ϵ(F) = 4.6) at the solar metallicity, somewhat higher than the Lodders et al. (2009) meteoritic abundance of log ϵ(F) = 4.42 or the new Maiorca et al. (2014) solar photospheric abundance of log ϵ(F) = 4.40." Pilachowski & Pace (2015) suggested that "the Sun may be slightly deficient in fluorine compared to the solar neighborhood, but a firm conclusion awaits a self-consistent determination of stellar atmospheric parameters using infrared spectra." Our revised CI-chondrite F-abundance is now more consistent with the K dwarfs and giants by Pilachowski & Pace (2015).

**Table 14. Present Day Fluorine Abundances**
**A(F) = log ϵ(F) = 12+ log (F/H)**

| | A(F) | ±dex | Reference |
|---|---|---|---|
| Meteoritic | 4.61 | 0.09 | (1) |
| Meteoritic | 4.44 | 0.22 | (2) |
| Sunspot | (4.56)[a] | | (3) |
| Sunspot | 4.40 | 0.25 | (4) |
| Sunspot | 4.37 | 0.04 | (5) |
| 67P/Churyumov-Gerasimenko | 4.63[b] | 0.66 | (6) |
| (from F/O ratio) | 4.67[c] | 0.66 | |
| 6 K dwarfs average | (4.60)[a] | 0.13 | (7) |
| 7 K dwarfs average | 4.63 | 0.25 | (8) |
| Fit for K giants in solar neighborhood at [Fe/H]=0 | 4.60 | | (8) |
| A(F) vs. A(O) for K giants in solar neighborhood | 4.37[b]<br>4.45[c] | | (9) |
| Pre-mainsequence stars in Orion | (4.60)[a] | 0.06 | (10) |
| Open Cluster M67 | 4.49 | 0.2 | (11) |
| Interstellar Medium | 4.17[d] | 0.17 | (12) |

[a] Older excitation potential for HF feature is used, see Pilachowski & Pace (2015) for discussion.
[b] for A(O) =8.69 from Allende-Prieto et al. (2001)
[c] for A(O) = 8.73, from Lodders et al. (2009)
[d] ISM toward two stars in the Cet OB2 association; using H from Cartledge et al. (2006)
References - (1) This work; (2) Palme et al. 2014; (3) Hall & Noyes 1969; (4) Maiorca et al. 2014; (5) Abia et al 2015; (6) Dhooghe et al. 2017, Rubin et al. 2019; (7) Recio-Blanco et al 2012; (8) Pilachowski & Pace 2015; (9) Jönsson et al. 2017; (10) Cunha & Smith 2005; (11) Maiorca et al. 2014 ; (12) Federman et al. 2005, Snow et al. 2007

Jönsson et al. (2017) measured F in K-giants in the solar neighborhood and found a significant correlation of fluorine and oxygen abundances which is understood by updated galactic chemical evolution models considering contributions to the fluorine inventory by AGB stars and Wolf Rayet star winds and novae (see Spitoni et al. 2018). The fit by Jönsson et al. (2017) yields A(F) = 4.37 for A(O) = 8.69 from Allende-Prieto et al. (2001); and A(F) = 4.45 with A(O) = 8.73 from Lodders et al. (2009). Both values are more consistent with sunspot values but lower than the CI-chondritic F abundance, or the stellar values by Pilachowski & Pace (2015). The only F-values in Figure 4 which do not overlap with CI-chondrites is the ISM fluorine abundance by Federman et al. (2005) and Snow et al. (2007), indicating a fluorine depletion in the ISM.

## 4.2 Chlorine

Table 15 and Figure 5 summarize chlorine abundances. The CI-chondritic Cl/$10^6$ Si ratio was scaled to the H-based abundance scale by using the atomic Si/H ratio of the photosphere (3.24e-5). The same conversion was done for the chlorine concentration (776 ppm) for asteroid Ryugu measured by Yokoyama et al. (2022). The chlorine concentration found for asteroid Ryugu is very similar to the selected values of CI-chondrites here. Yokoyama et al. (2022) found that Ryugo has quite similar compositions of the elements as seen in CI-chondrites. Their finding of a high chlorine abundance strengthens the argument that the high chlorine concentration seen in CI-chondrites is real and not due to contamination of CI-chondrites as argued by one reviewer.

The solar photosphere offers no useful atomic lines to determine the chlorine abundances. Lambert et al. (1971) could only derive an upper limit of A(Cl) < 5.5 from a weak atomic line in the photospheric infrared spectrum. The cooler sunspot umbra allows the determination of the F and Cl abundances from their hydrogen halide rotation-vibration-spectra. Hall and Noyes (1972) found A(Cl) = 5.5± 0.3. Maas et al. (2016) found A($^{35}$Cl) = 5.31±0.12 from their best fit of a sunspot spectrum. Assuming the same (first order) ratio of $^{35}$Cl/$^{37}$Cl = 3.127 as found for Earth and meteorites, the total abundance from sunspots is A(Cl) = 5.43±0.12. Another direct chlorine measurement for the sun gives A(Cl) = 5.75±0.26 from X-ray spectra of highly ionized Cl in solar flares (Sylwester et al. 2011). Solar wind as measured from solar energetic particles (SEP) gives A(Cl) = 5.70±0.15 (Reames 2018) where uncertainties in the H-abundance and the fractionation between H and other elements with low ionization potentials also affects the results. The gas phase Cl/O ratios observed in comet 67P/Churyumov-Gerasimenko range from 2 $\times 10^{-4}$ to 5 $\times 10^{-4}$ with a weighted average Cl/O = 1.19 $\times 10^{-4}$ (Dhooghe et al. 2017, Rubin et al. 2019; note that their weighted average is lower than the range quoted in Table 3 of Dhooghe et al. 2017). The nominal abundances (with considerable uncertainties) are among the lowest values in Table 15 and Figure 5 and may suggest that chlorine was not fully outgassed in the comet.

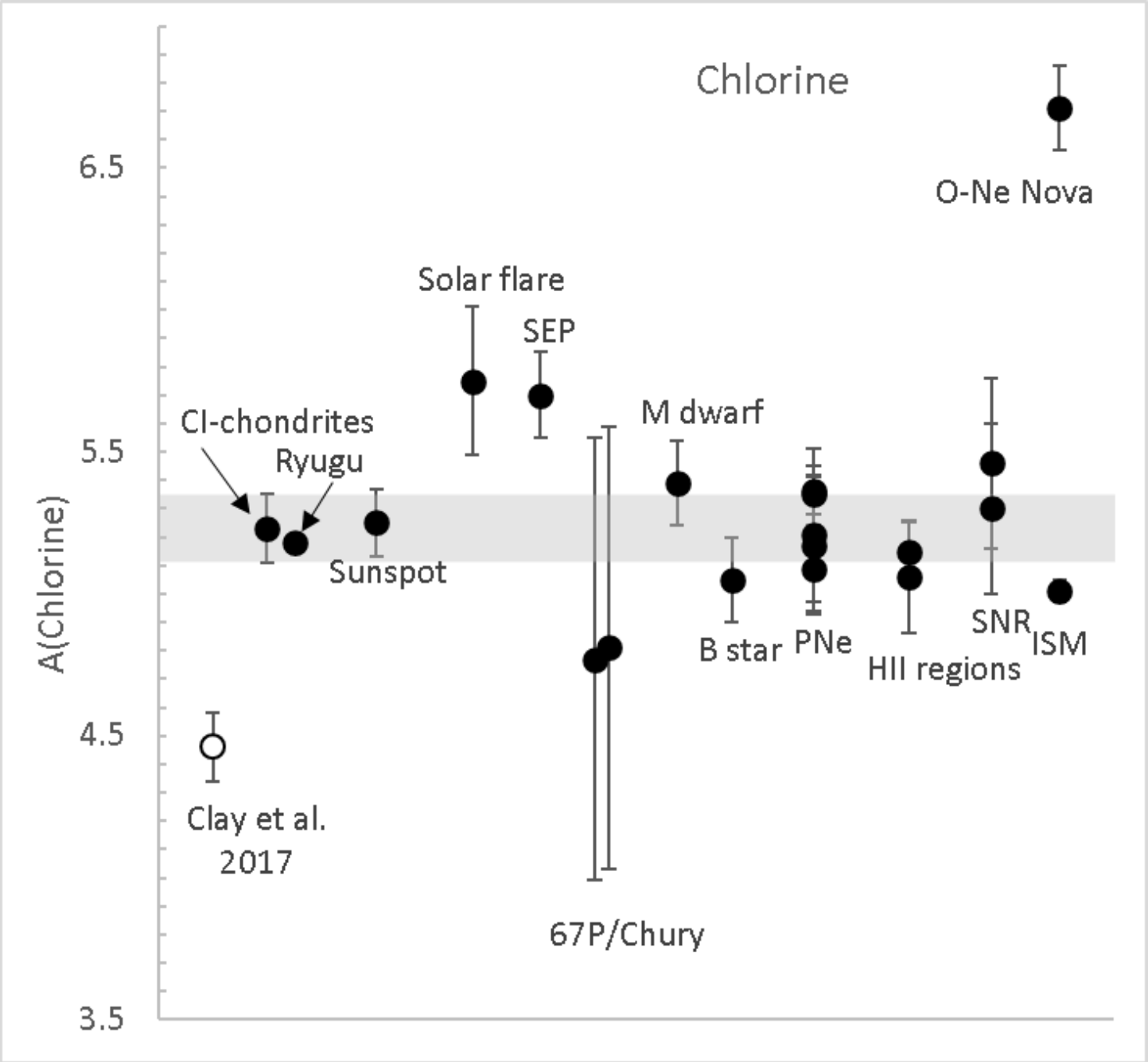

**Figure 5**. Chlorine abundances in CI-chondrites, sunspots, and other objects as indicated. The supernova remnant and comet values are re-scaled to the photospheric H abundance for comparison, see text. The shaded band is the range for CI-chondrites. For data sources see Table 15. If no uncertainties were given, an uncertainty of 0.15 dex was assumed.

The lowest Cl abundance shown in Figure 5 (open symbol) from the Cl-chondrite measurement by Clay et al. (2017); clearly illustrate that their analyses have problems. A representative solar chlorine value should at least be in the same order of magnitude as that measured for the sun and other stars.

Maas et al. (2016) determined chlorine abundances in 15 cool K-M giant stars and one M-dwarf star from the $^{35}$Cl-H rotation-vibration-lines in the L-band spectra. They could only obtain abundances for the more-abundant $^{35}$Cl isotope, which is about three times more abundant than $^{37}$Cl in the solar system. The M-dwarf star BD+68 946 (M3.5V) has solar metallicity [Fe/H] = 0, and A($^{35}$Cl) = 5.27, which, assuming terrestrial isotopic composition, gives A(Cl) = 5.39, consistent with solar values within uncertainties.

Mass et al. (2016) included K and M giants with different metallicity and at different stages of evolution; several known S and Ba stars are included. The Ba and S stars became enriched in carbon (from triple-Helium-burning) and elements produced by the slow neutron-capture process (s-process) when these stars evolved to the AGB stage and experienced multiple "3$^{rd}$ dredge ups" that brought processed material to their stellar surfaces; or by mass transfer from an evolved companion (extrinsic Ba and S stars). If there were any contributions to the galactic chlorine budget from AGB nucleosynthesis one should expect Cl enrichments in these objects. The Ba stars measured by Maas et al. have similar $^{35}$Cl abundances for given metallicity as normal M and K giants, and larger contribution from the s-process to the Cl abundance are unlikely. A similar conclusion is reached from observations of planetary nebulae (descendants of AGB stars) which have similar relative abundances of Cl as other light elements not made by the s-process (e.g., Henry et al. 2004, Hyuang & Feibelman 2004, Madonna et al. 2017).

**Table 15. Present Day Chlorine Abundances**
**A(Cl) = 12+ log (Cl/H)**

| | A(Cl) | ± dex | Reference |
|---|---|---|---|
| Meteoritic, CI-chondrites | 5.23 | 0.06 | (1) |
| Ryugu Asteroid (akin to CI chon.) | 5.18 | 0.02 | (2) |
| Photosphere | < 5.5 | | (3) |
| Sunspot | (5.5)[a] | 0.3 | (4) |
| Sunspot | 5.25 | 0.12 | (5) |
| Solar flare | 5.75 | 0.26 | (6) |
| Solar Energetic Particles | 5.70 | 0.15 | (7) |
| 67P/Churyumov-Gerasimenko[b,c] | 4.77[b] | 0.78 | (8) |
| | 4.81[c] | 0.78 | |
| M Dwarf BD+68 0946 | 5.39 | | (9) |
| B star $\pi$ Cet | 5.05: | *uncert.* | (10) |
| Planetary Nebulae [b,c] | 5.17[b] | 0.24 | (11) |
| | 5.21[c] | 0.24 | |
| Planetary Nebula NGC 6803 | 5.36 | | (12) |
| Planetary Nebula NGC 5315 | 5.35 | 0.07 | (13) |
| Planetary Nebulae at 8 kpc | 5.09 | | (14) |
| HII Regions within 2 kpc of Sun | 5.15 | 0.1 | (15) |
| HII Regions at 8 kpc | 5.06 | 0.2 | (16) |
| Supernova Remnant [d,e] | 5.46[d] | 0.3 | (17) |
| | 5.30[e] | 0.3 | |
| Interstellar medium | 5.01 | 0.04 | (187) |
| Nova V1974 Cyg | 6.71 | | (19) |

[a] Superseded by more recent analyses.
[b] from C/O for A(O) =8.69 from Allende-Prieto et al. (2001)
[c] from Cl/O for A(O) = 8.73 from Lodders et al. (2009)
[d] from Cl/S = 0.020 scaled to photospheric A(S) = 7.16 (Palme et al. 2014)
[e] from Cl/Ar = 0.063 scaled to photospheric A(Ar) = 6.50 (Lodders 2008)
References - (1) This work; (2) Yokoyama et al. 2022; (3) Lambert 1971; (4) Hall & Noyes 1972; (5) Maas et al. 2016 ; (6) Sylwester et al. 2011; (7) Reames 2019; (8) Rubin et al. 2019, Dhooghe et al. 2017; (9) Mass et al. 2016; (10) Fossati et al. 2009; (11) Henry et al. (2004); (12) Hyung & Feibelman 2004; (13) Madonna et al. 2017; (14) Henry et al. 2004; (15) Rodriguez 1999; (16) Esteban et al. 2015; (17) Seitenzahl et al. 2012; (18) Moomey et al. 2012; (19) Arkhipova et al. 1997

Rodriguez (1999) found an average A(Cl) = 5.15±0.10 for Galactic HII regions within 2 kpc of the sun. The abundances from emission lines of highly ionized Cl in HII-regions shows an enrichment toward the Galactic Centre as a function of radial distance (Rg in kpc). Esteban et al. (2015) give the following fits from their observations for chlorine:

$$A(Cl) = 12 + \log(Cl/H) = 5.40(\pm 0.11) - 0.043(\pm 0.012) \times Rg$$

Adopting Rg = 8 kpc for the Sun gives A(Cl) = 5.06 for the solar neighborhood from measurements of HII regions (Esteban et al. 2015); using a similar relation for planetary nebulae from Henry et al. (2004) gives A(Cl) = 5.09, both lower than in sunspots and our recommended CI-chondritic chlorine values (Table 15). If these differences are not caused by uncertainties in the atomic parameters used in the analyses, the lower Cl abundances measured in the gas phases of the HII regions and PNe suggests a depletion of ionized Cl. This could be due to Cl depletion into grains, or that Cl occurs as NaCl gas and other halogen gases such as AlCl and KCl which have been detected in circumstellar shells of AGB stars (Cernicharo & Guelin, 1987, Milam et al. 2007, Decin et al. 2018, de Beck & Olofsson 2018) and suggests that NaCl gas is a chlorine host in the ISM.

The Cl abundance determined in a supernova remnant by Seitenzahl et al. (2012) was calculated from the Cl/S and Cl/Ar ratios assuming photospheric S and Ar abundances because H is depleted in the SNR and the reported abundances relative to H are thus too high for comparison with solar values. We used S and Ar for scaling because chlorine isotopes are mainly produced together with major isotopes of the neighboring elements S and Ar. The two isotopes $^{35}$Cl and $^{37}$Cl are mainly produced during hydrostatic oxygen and neon burning in massive stars, or during explosive oxygen burning in core-collapse supernovae type SNII, and supernovae type Ia (e.g., Travaglio et al. 2004, Kobayashi et al. 2006, Nomoto et al. 2013; Maas et al. 2016 give a detailed discussion of Galactic chemical evolution for chlorine and its isotopes). Thus, the relative abundances of S, Cl, and Ar should be very similar in various astronomical environments unless chemical volatility fractionations, or physical fractionations due to electromagnetic plasma interactions affected observed abundances. The more refractory S than Cl can be depleted in the gas phase by condensation into sulfides; or elements with different ionization potentials can fractionate in solar or stellar plasma environments. Within uncertainties, the scaled SNR chlorine abundances in Figure 5 agree and are a little higher than the CI-chondritic value. The Cl/S ratio of the SNR yields a slightly higher Cl abundance which could indicate that some more refractory sulfur was lost into grains.

The other exception in Table 15 and Figure 5 is the higher Cl abundance for the O-Ne nova by Arkhipova et al. (1997). We did not re-scale the reported abundance to the photospheric H abundance as done for the SNR (rescaling with Cl/Ar and solar Ar would reduce the Cl abundance by 0.3 dex only and still leave it more than 1 dex above the sunspot value). This object has a Cl/Ar ratio about a factor of ten higher than the sun, PNe, and the SNR indicating a genuine Cl-enrichment. A relative enrichment of Cl in novae nucleosynthesis is plausible (see Jose et al. 2004), but more measurements on O-Ne novae need to confirm this.

### 4.3 Bromine and Iodine

There are only a few comparisons of the meteoritic Br abundance to other astronomical objects, and none for I. The CI-chondritic Br abundance scaled to photospheric hydrogen is A(Br) = 12 + log (Br/H) = 2.60±0.09; and for iodine we obtain A(I) = 12 + log (I/H) = 1.71 ± 0.15. There are no Br abundance measurements for the Sun and other dwarf stars. The gas phase Br/O in comet 67P/Churyumov-Gerasimenko ranges from (1-7) $\times 10^{-6}$ with a weighted average Br/O ~ 2.48(+4.46; -1.49) $\times 10^{-6}$ (Dhooghe et al. 2017, Rubin et al. 2019). This corresponds to A(Br) = 3.08±0.45 using photospheric A(O)= 8.69, and 3.12±0.45 for A(O) =8.73. Within uncertainties, the meteoritic and cometary values overlap; however, the nominal cometary Br abundance is higher than the meteoritic value when both are scaled to the photospheric H abundance.

Madonna et al. (2017) analyzed lines of $Br^{2+}$ in the planetary nebula NGC 5315 and derived an approximate Br abundance of A(Br) =3.53 from a possibly blended line, and an upper limit A(Br) < 2.43 from another line. Future analyses are needed to obtain independent constraints on the Br and I abundances from stellar sources.

## 4.4 Halogen Abundance Recommendations: Meteoritic, Solar, and Solar System

Table 16 summarizes our recommended halogen abundances for CI-chondrites, solar, and the solar system. The two atomic abundance scales (see footnote Table 16) are linked via $A(X) = 1.51 + \log_{10} N(X)$ and relate to the present-day heavy element to hydrogen ratios in the upper solar atmosphere. Over the sun's lifetime, element diffusion and settling from the outer convection zone decreased the element-to-hydrogen ratios. Therefore, a setting correction is made to obtain the element/hydrogen ratios that the sun had at its birth; these are the solar system or proto-solar abundances 4.567 Ga ago. The settling correction is usually assumed to be the same factor for elements heavier than helium, so there is no change in the elemental abundances on the atomic scale relative to silicon. The settling factor to increase the X/H ratios to protosolar values (or the constant for the log H = 12 scale) depends on solar models and constraints from helioseismology (e.g., Lodders 2003, 2020) and values range from 12-22% (0.05 – 0.088 dex). Here we use our currently preferred factor of 0.088 dex based on the models by Yang (2019) and obtain the solar system values from $A(X)_0 = 1.51 + \log_{10} N(X) + 0.088$. If another settling factor is preferred, the proto-solar values are easily computed from the present-day values.

**Table 16. Recommended Abundances of the Halogens**

| | CI-Chondrites | Solar: CI-Chondrites | | Solar: Sunspots | | Solar System [c] | |
|---|---|---|---|---|---|---|---|
| | ppm by mass | N(X) [a] | A(X) [b] | N(X) [a] | A(X) [b] | $N(X)_0$ [a] | A(X)0 b |
| F | 92 ± 20 | 1270 ± 275 | 4.61 ± 0.09 | 780 ± 600 | 4.40 ± 0.25 | 1270 ± 275 | 4.70 ± 0.09 |
| Cl | 712 ± 110 | 5290 ± 810 | 5.23 ± 0.06 | 5500 ± 1750 | 5.25 ± 0.12 | 5290 ± 810 | 5.32 ± 0.06 |
| Br | 3.77 ± 0.90 | 12.3 ± 2.9 | 2.60 ± 0.09 | - | - | 12.3 ± 2.9 | 2.69 ± 0.09 |
| I | 0.77 ± 0.31 | 1.59 ± 0.64 | 1.71 ± 0.15 | - | - | 1.59 ± 0.64 | 1.80 ± 0.15 |

[a] N(X) on a scale normalized to Si = $10^6$ atoms.
[b] $A(X) = 12 + \log_{10}(N(X)/N(H))$.
[c] Proto-solar values; a settling correction of 0.088dex was applied to the logarithmic abundance scale, see text.

The recent measurement by Yokoyama et al. (2022) of 776± ppm chlorine (corresponding to A(Cl) = 5.18±0.02) in the returned samples from the CI-chondrite-like asteroid Ryugu is quite similar to our recommended value derived from CI chondrites in original submission of our paper. The higher chlorine concentration in Ryugo supports our notion that the high chlorine concentrations in CI chondrites measured since 1864 are intrinsic and not due to terrestrial contamination (see section 4.2). We note that there are no measurements for F, Br and I yet for Ryugu as the time of writing, and that further measurements on Ryugu and meteorite samples are required to obtain a better handle on the halogen abundances.

# 5. Condensation Temperatures of the Halogens

Halogen condensation temperatures are more variable than those of many elements because of uncertainties in halogen elemental abundances and thermodynamic data for possible halogen-bearing condensates. The condensation temperatures of the heavier halogens are relatively low and kinetic effects can become important if solid-gas reactions are involved. Diffusion in solid phases can become sluggish and equilibrium may not be attained during the time where nebular gas could react with preexisting condensates. We describe kinetic effects later in this section.

Over time updates in elemental abundances and/or thermodynamic data have led to improvements for the condensation calculations but uncertainties remain. For example, three different sets of thermodynamic data are available for the chlorine-bearing mineral sodalite (Wellman 1969, Komada et al 1995, Schliesser et al. 2017) and further work is needed to determine the "correct" thermodynamic properties of sodalite ($Na_4[Al_3Si_3O_{12}]Cl$). Condensation computations for one or more halogens were done by our group for F, Cl, Br (Fegley and Lewis 1980), F, Cl, Br, I (Lodders 2003), F, Cl, Br (Fegley and Schaefer 2010), and F, Cl, Br, I (Fegley and Lodders 2018), which presents the preliminary results of this work. Other halogen condensation calculations in the literature are for F (Wai and Wasson 1977), Cl (Sharp et al. 1989), and F, Cl, Br, I (Wood et al. 2019). However, the latter group used the Clay et al. (2017) halogen abundances, which are too low as discussed earlier.

Lodders (2003) describes the CONDOR code, computational methods and thermodynamic data sources in more detail than presented here. The condensation temperature of a pure mineral is the temperature at which the thermodynamic activity of the mineral is unity. The code computes thermodynamic activities iteratively subject to mass balance and mass action, i.e., equilibrium constant $K_{eq} = \exp(-\Delta G°/RT)$, constraints. The 50% condensation temperature of an element (mineral) that dissolves in a more abundant host phase depends upon the activity of the element (mineral) in the solution and the relevant activity coefficient ($\gamma = 1$ if ideal, $\gamma \neq 1$ if nonideal) because the mole fraction of the dissolved element (mineral) equals activity/activity coefficient. Abundance of the host phase times the mole fraction of the dissolved element (mineral) gives the fractional abundance of the element (mineral) condensed.

The differences between our present and earlier work are from the different solar abundances and different thermodynamic data used in the calculations. The present work used updated thermodynamic data compilations, which postdate Fegley and Lewis (1980), e.g. the 4$^{th}$ edition of the JANAF Tables (Chase 1998), Robie and Hemingway (1995). We use more recent papers giving updated thermodynamic properties for apatites, calcium phosphates, sodalite, schreibersite ($Fe_3P$), several P- and S-bearing gases, and the OH gas dissociation energy (Cruz et al. 2005, Dachs et al. 2010, Drouet 2015, Komada et al. 1995, Lodders 1999a, 2004, Ruscic et al 2014, Schliesser et al. 2017, Zaitsev et al. 1995). Some incorrect data are in all editions of the JANAF Tables and we make corrections where necessary, e.g. the OH dissociation energy, and data for several P- and S-bearing gases (Lodders 1999a, 2004, Ruscic et al. 2014).

## 5.1. Recommended halogen and phosphorus condensation temperatures

Our recommended 50% equilibrium condensation temperatures at $10^{-4}$ bars total pressure in the solar nebula are 1264 K for phosphorus in schreibersite $(Fe,Ni)_3P$, 713 K for fluorine in fluorapatite $Ca_5(PO_4)_3F$, 427 K for chlorine in halite NaCl, 392 K for bromine in sodium bromide NaBr, and 312 K for iodine in sodium iodide (Table 17). We discuss possible kinetic inhibition of apatite and sodalite condensation in sections 5.2 and 5.3. Our recommended values for solar abundances and 50% equilibrium condensation temperatures for F, Cl, Br and I agree better with the lithophile volatility trend for chondrites and Earth than previously (see Fegley et al. 2020).

## 5.2. Fluorine and chlorine condensation chemistry

Fluorapatite starts condensing at 717 K and remains stable to 298 K where we stopped calculations. Chlorapatite $Ca_5(PO_4)_3Cl$ starts condensing at 648 K and forms via oxidation of schreibersite. Chlorapatite stops forming at 632 K because its formation consumes all available phosphorus in schreibersite. Chlorapatite remains stable to 298 K. Chlorine remains in the gas (as mainly HCl + NaCl) until 430 K where halite NaCl starts forming. Halite is present from 430 - 408 K where sodalite becomes stable. Sodalite is stable from 408 - 369 K where halite again becomes stable to 298 K. Table 17 summarizes this set of reactions and those for Br and I discussed below.

Fegley and Schaefer (2010) described the mass-balance involved with F and Cl condensation. We go through their arguments using the updated halogen abundances derived earlier in this work. Fluorine is the least abundant element in fluorapatite and its abundance limits that of fluorapatite. Fluorapatite has a 3:1 ratio of P to F atoms and the CI-chondritic P:F atomic ratio is 8350:1270. The complete condensation of fluorine into fluorapatite consumes about 46% of total phosphorus leaving 54% of total phosphorus for chlorapatite formation. But the remaining amount of phosphorus is insufficient to condense all chlorine into chlorapatite, which has a 3:1 atomic ratio of P to Cl atoms. After fluorine condensation and before chlorine condensation, the remaining P: Cl atomic ratio is 4540:5290 and the maximum amount of Cl that can condense into Cl-apatite is 4540/3 = 1513 atoms (29% of total chlorine). The residual chlorine condenses as halite NaCl, which appears at 430 K, and the chlorine 50% condensation temperature is 427 K with 29% of chlorine in chlorapatite and 21% in halite.

**Table 17. Equilibrium Halogen Condensation Temperatures at $10^{-4}$ bar from Solar Gas[a]**

| | $T_C$ start | Condensate | 50% $T_C$ | Notes |
|---|---|---|---|---|
| F | 717 | $Ca_5(PO_4)_3F$, Fluorapatite | 713 | HF is major F gas |
| Cl | 648 | $Ca_5(PO_4)_3Cl$, Chlorapatite | | 29% of Cl in Cl-apatite, mass balance |
| | 430 | NaCl, Halite | 427 | NaCl 430-408 K |
| | 408 | $Na_4(AlSiO_4)_3Cl$, Sodalite | | Sodalite 408-369 K |
| | 369 | NaCl, Halite | | Halite stable again 369-300 K |
| Br[b] | 397 | NaBr[b] | 392 | HBr is major Br gas |
| I[c] | 321 | NaI[c] | 312 | HI is major I gas |

[a] Solar abundances from Table 1.2 of Lodders & Fegley (2011) and the halogen abundances recommended in this work.
[b] For pure NaBr; activity coefficient for NaBr in NaCl ~3.5 in 430 – 400 K range, 50% of bromine condensed at 430 K in NaCl.
[c] For pure NaI; activity coefficient for NaI in NaCl ~ 370 – 760 in 430 – 380 K range; 50% of iodine is condensed at 381 K in NaCl.

## 5.3. Bromine and iodine condensation chemistry

Bromine remains in the gas as HBr until sodium bromide condenses at 397 K. Iodine remains in the gas as HI until sodium iodide condenses at 321 K. Bromine and iodine condensation are slightly different for fast and slow cooling of nebular gas (cf. Larimer and Anders 1967). Rapid cooling gives pure NaBr and NaI and one can imagine reactions such as those below taking place.

$$\text{HBr (gas)} + \text{NaCl (halite)} = \text{NaBr (crystal)} + \text{HCl (gas)}$$

$$\text{HI (gas)} + \text{NaCl (halite)} = \text{NaI (crystal)} + \text{HCl (gas)}$$

The HCl expelled into the gas could condense as $NH_4Cl$ and/or $HCl \cdot 3H_2O$ at lower temperature (Fegley and Lewis 1980, Zolotov and Mironenko 2007). Ammonium chloride was observed in the Orgueil CI-chondrite shortly after its fall (Cloëz 1864, Pisani 1864). More recently, Poch et al (2020) observed ammonium salts, possibly including ammonium halides, on comet 67P/Churyumov-Gerasimenko.

Slow cooling allows solid state diffusion and then NaBr and NaI may form solid solutions with halite. These solid solutions are energetically unfavorable because bromide (1.96 Å radius) and iodide (2.20 Å radius) ions are larger than chloride ions (1.81 Å radius). The activity coefficients for NaBr or NaI dissolution in NaCl are ~3.5 in the 430 - 400 K range for NaBr and ~370 - 760 in the 430 - 380 K range for NaI (Sangster and Pelton 1987). Formation of the solid solutions increases the 50% condensation temperatures to 430 K for bromine and 381 K for iodine. Condensation of $NH_4Cl$ and/or $HCl \cdot 3H_2O$ could also occur if NaBr-NaCl and NaI-NaCl solid solutions formed.

Fegley and Lewis (1980) found bromine condensed as pure bromapatite $Ca_5(PO_4)_3Br$ at about 350 K on the solar nebula adiabat using the standard Gibbs energy of formation $\Delta_f G°$ at 298 K from Duff (1972) and estimated entropy and heat capacity data. However, the standard enthalpy of formation $\Delta_f H°$ at 298 K measured by Cruz et al (2005) makes bromapatite much less stable. But the absolute entropy at 298 K and heat capacity of bromapatite are unknown. These data are needed for accurate assessment of bromapatite stability. There are no measured entropy, heat capacity, $\Delta_f H°$ or $\Delta_f G°$ data for iodapatite $Ca_5(PO_4)_3I$ and its stability cannot be computed accurately. We think its formation is unlikely because of the high volatility of iodine in solar nebula gas and the probable low temperatures required for its formation. In absence of complete thermodynamic data for Br- and I-apatite, Lodders (2003) assumed substitutions of the Ca-fluoride and Ca-chloride component in apatite by the respective Ca-bromide and iodides. This gave 50% condensation temperatures of ~546 K (Br) and 535 (I). Condensation calculations for ideal dissolution of $FeBr_2$ or $FeI_2$ into troilite FeS gave negligible amounts of either compound.

### 5.4. Gas-solid reactions involved in halogen condensation

In the following sections we discuss specific gas - solid reactions that may be involved in halogen condensation. Fluorine and chlorine condense via reaction of preexisting condensates with nebular gas at lower temperatures. As seen below the net chemical reactions are complicated and involve several different condensates that must equilibrate with one another. It is uncertain whether the condensation reactions reach equilibrium during the lifetime of the solar nebula, but quantitative consideration of this question is beyond the scope of this paper. There are also differences in detail between the predicted and observed composition of apatite in chondrites that lead us to question the canonical condensation sequence for halogens.

We emphasize our discussion of specific reactions is for pedagogical purposes only. Chemical equilibrium codes – unlike chemical kinetic codes – do *not* require calculations based upon specific reactions because the Gibbs free energy is a state variable and is path independent.

## 5.5. Fluorine and chlorine condensation into fluor- and chlorapatite

Fluor- and chlorapatite contain calcium and phosphorus, which are completely condensed hundreds of degrees higher than fluor- or chlorapatite become stable. Furthermore, calcium is more refractory than phosphorus (50% condensation temperatures at $10^{-4}$ bars total pressure of 1517 K for Ca and 1265 K for P) and initially condenses as minerals such as hibonite ($CaAl_{12}O_{19}$), and gehlenite ($Ca_2Al_2SiO_7$), which occur in Ca, Al-rich inclusions (CAI) in meteorites (Lodders 2003). Phosphorus condenses into an iron, nickel phosphide called schreibersite $(Fe,Ni)_3P$, a common meteoritic mineral. The good fit of the phosphorus 50% condensation temperature on the volatility trend for CM and CV chondrites and the rapid corrosion of $\alpha$-Fe metal by phosphorus vapor suggest P condensation into schreibersite reached chemical equilibrium (Fegley et al. 2020, Sasaki and Ueda 1972). The formation of $(Fe,Ni)_3P$ by corrosion of (Fe,Ni) alloy by $P_2$, PO, PS, P, and PH in the solar nebula (Fegley and Schaefer 2010) involves a volume change of about 30% per mole schreibersite formed. This large $\Delta V$ is likely to crack phosphide layers on metal, exposing fresh metal to gas, and give faster reaction than suggested by diffusion-controlled kinetics.

Fluorapatite condensation involves oxidation of $(FeNi)_3P$, which may involve the intermediate formation of the H-free phosphate merrillite $Ca_9Na(Mg, Fe^{2+})(PO_4)_7$, depending on the poorly known sequence of thermodynamic stability of merrillite and whitlockite. The mineral name whitlockite is now reserved for the hydrous phosphate, $Ca_9Mg(PO_4)_6(PO_4H)$ which is not common to chondrites; however, merrillite and whitlockite are still used interchangeably in the cosmochemical literature. Merrillite and apatite often occur in association with olivine, pyroxenes, feldspar, metal and sulfides in ordinary chondrites (see Lewis and Jones 2016 for a detailed study), and like apatite, merrillite requires a calcium source for its formation. This source could be one or more refractory minerals which we denote as "CaO (in condensate)". Following prior work, we use the simplified formula $Ca_3(PO_4)_2$ for the phosphate (Wai and Wasson 1977, Sears 1978, Fegley and Lewis 1980). Possible exemplary reactions for fluorapatite formation from pre-existing phosphide via phosphate formation include the sequence

$$\begin{aligned}&3\mathrm{CaO(in\ condensate)} + 2\mathrm{(Fe,Ni)_3P(schreibersite)} + 5\mathrm{H_2O(gas)}\\&\quad = \mathrm{Ca_3(PO_4)_2(merrillite)} + 6\mathrm{(Fe,Ni)(alloy)} + 5\mathrm{H_2(gas)}\end{aligned}$$

followed by:

$$\begin{aligned}&\frac{3}{2}\mathrm{Ca_3(PO_4)_2(merrillite)} + \frac{1}{2}\mathrm{CaO(in\ condensate)} + \mathrm{H(F,Cl)\ (gas)}\\&\quad = \mathrm{Ca_5(PO_4)_3(F,Cl)[fluor(chlor)apatite]} + \frac{1}{2}\mathrm{H_2O\ (gas)}\end{aligned}$$

These reactions involve either HF (forming fluorapatite) or HCl (forming chlorapatite), which we abbreviate as H(F,Cl). Wai and Wasson (1977) discussed another possible reaction,

$$\begin{aligned}&\mathrm{H(F,Cl)\ (gas)} + \frac{3}{2}\mathrm{Ca_3(PO_4)_2(merrillite)} + \frac{1}{2}\mathrm{CaMgSi_2O_6(diopside)} + \frac{1}{2}\mathrm{Mg_2SiO_4(forsterite)}\\&\quad = \mathrm{Ca_5(PO_4)_3(F,Cl)[fluor(chlor)apatite]} + \frac{3}{2}\mathrm{MgSiO_3(enstatite)} + \frac{1}{2}\mathrm{H_2O\ (gas)}\end{aligned}$$

Fegley and Schaefer (2010) found that fluorapatite forms at 710 K from schreibersite; our calculations with updated elemental abundances and thermodynamic data revise this to 717 K (Table 17). Examples of net thermochemical reactions for apatite formation directly from schreibersite are:

$$5\mathrm{CaO(in\ condensate)} + 3\mathrm{(Fe{,}Ni)_3P(schreibersite)} + 7\mathrm{H_2O(gas)} + \mathrm{HF\ (gas)}$$
$$= \mathrm{Ca_5(PO_4)_3F\ (fluorapatite)} + 9\mathrm{(Fe{,}Ni)(alloy)} + \frac{15}{2}\mathrm{H_2(gas)}$$

Or

$$5\mathrm{CaMgSi_2O_6(diopside)} + 3\mathrm{(Fe{,}Ni)_3P(schreibersite)} + 5\mathrm{Mg_2SiO_4(forsterite)} + \mathrm{HF(gas)} + 7\mathrm{H_2O(gas)}$$
$$= \mathrm{Ca_5(PO_4)_3F\ (fluorapatite)} + 9\mathrm{(Fe{,}Ni)(alloy)} + 15\mathrm{MgSiO_3(enstatite)} + \frac{15}{2}\mathrm{H_2(gas)}$$

All these reactions involve several minerals and solid-state diffusion of elements between the different minerals. Chemical kinetic modeling of apatite formation reactions is beyond the scope of this paper. Volume diffusion of Ca, Fe, Na, P, and other cations is likely to be slower than either grain boundary and surface diffusion at the calculated equilibrium temperatures for apatite (and sodalite) formation, and all diffusion pathways may be too slow to be important on solar nebula timescales (e.g., see Hofmann et al 1974, Freer 1981). But in the absence of experimental studies of the proposed reactions, we cannot say more about diffusion. The good agreement of the 50% condensation temperature for fluorine with the volatility trend for chondrites and Earth suggests an alternative reaction such as HF corrosion of silicate minerals and/or metal alloy giving metal fluorides ($MgF_2$, $CaF_2$, $FeF_2$) plus water vapor may be responsible for fluorine condensation. One possible schematic reaction is

$$2\mathrm{HF\ (gas)} + \mathrm{CaO\ (in\ condensate)} = \mathrm{CaF_2\ (fluorite)} + \mathrm{H_2O\ (gas)}$$

Subsequent metamorphic reactions on meteorite parent bodies may have formed F-bearing apatites. We think the absence of halogen-bearing apatite minerals in the lowest metamorphic grade chondrites and the presence of halogen-bearing apatite minerals in higher metamorphic grade chondrites tells us that meteoritic apatite formed on parent bodies (see section 5.3).

### 5.2.2 Sodalite

Chlorine is part of the mineral sodalite, $Na_4[Al_3Si_3O_{12}]Cl$ = "$3NaAlSiO_4 \cdot NaCl$". Fegley & Lewis (1980) and Lodders (2003) used thermodynamic data from Wellman (1969) for sodalite condensation calculations. Fegley & Lewis (1980) found sodalite condensed at 895 K and chlorine was 50% condensed in sodalite at 873 K along an adiabatic P, T profile for the solar nebula (Lewis 1974). Lodders (2003) calculated sodalite condensed at 954 K and 50% of Cl condensed at 948 K into sodalite at $10^{-4}$ bar total pressure (cf. Fegley & Schaefer 2010).

Use of the new data shows sodalite is less stable than calculated previously (Fegley & Lewis 1980, Lodders 2003), and it is not the first Cl-bearing condensate from a solar composition gas at $10^{-4}$ bar total pressure. The formation of sodalite involves chlorination of high-temperature condensates such as feldspar and nepheline, and occurs via the reaction (Ikeda & Kimura 1996, Kimura & Ikeda 1995)

$$3\ \mathrm{NaAlSi_3O_8\ (albite\ dissolved\ in\ feldspar)} + \mathrm{NaCl\ (gas)} = \mathrm{Na_4[AlSiO_4]_3Cl\ (sodalite)} + 6\ \mathrm{SiO_2\ (silica)}$$

Pure silica does not form but the coexistence of enstatite ($MgSiO_3$) and forsterite ($Mg_2SiO_4$) regulates the silica thermodynamic activity via the reaction (Carmichael et al. 1970),

$$2MgSiO_3 \text{ (enstatite)} = Mg_2SiO_4 \text{ (forsterite)} + SiO_2 \text{ (silica)}$$

Another possible set of sodalite - forming reactions is

$$CaSi_2Al_2O_8 \text{ (anorthite in feldspar)} + 2\ Na \text{ (gas)} + H_2O \text{ (gas)} = Na_2Si_2Al_2O_8 \text{ (nepheline)} + H_2 \text{ (gas)}$$

$$3Na_2Si_2Al_2O_8 \text{ (nepheline)} + 2\ NaCl \text{ (gas)} = 2\ Na_4[AlSiO_4]_3Cl \text{ (sodalite)}$$

Other net thermochemical reactions with HCl (gas) are possible. Ikeda & Kimura (1996) experimentally studied sodalite-forming reactions to study alkali and chlorine uptake by chondrules of the Allende CV3 chondrite.

## 5.3 Possible kinetic inhibition of apatite and sodalite formation

As described in section 2, halogen host phases in meteorites are alkali halide salts, fluor- and chlorapatite, and sodalite as noted above. However, the reactions involved in apatite and sodalite formation are complex and involve diffusion of several cations and anions among different minerals. The necessity to bring all solid phases together for apatite (or sodalite) formation is more plausible to occur on the meteorite parent body than in the solar nebula itself. For example, during thermal metamorphism on a meteorite parent body, liquid water (instead of water vapor) can transport halogens.

Thermal metamorphism and aqueous alteration on meteorite parent bodies caused re-distribution of the halogens from their primary condensates into the secondary mineral host phases observed in meteorites. The most important secondary phases are apatite and sodalite (see Brearley & Jones 2018). A good example is the study by Miura et al (2014), who measured noble gases in Allende (CV3) chondrules and found correlated enrichments in noble gases produced from halogens by neutron-capture reactions ($^{36}Ar$ from $^{35}Cl$, $^{80}Kr$ from $^{79}Br$, $^{82}Kr$ from $^{81}Br$, and $^{128}Xe$ from $^{127}I$, similar to the reactions used for halogen analyses described in section 2) on the meteorite parent asteroid. This occurred after chondrules had experienced dry alteration leading to nepheline and sodalite formation, see also Kimura & Ikeda (1995). Thus, secondary sodalite is the major host for chlorine, and apparently also for Br and iodine. This is similar to the halogen-bearing apatite, which is not a primary nebular condensate or circumstellar phase but instead a phase produced during thermal metamorphism in chondrites (e.g., Anders 1988, Lewis & Jones 2016, Brearley & Jones 2018, Ward et al. 2017). Indicators for a secondary origin of apatite and other phosphates are the increasing occurrence of apatite in chondrites with metamorphic equilibration grade (apatite is quite rare in unequilibrated ordinary chondrites (see Lewis & Jones 2016; Brearley & Jones 2018), and its rare earth element (REE) concentrations (e.g., Ward et al. 2017). The REE condense into perovskite and/or hibonite in CAIs (Kornacki & Fegley 1986). However, phosphate is the host for REE in most chondrites. The REE moved into phosphate minerals during metamorphic reactions on meteorite parent bodies. As noted already, apatite in chondrites is generally Cl-rich and F-rich apatite is extremely rare. The reverse situation occurs in the equilibrium condensation sequence: F-rich apatite is thermodynamically more stable and condenses at higher temperature before Cl-rich apatite forms.

## 5.4 Kinetic inhibition condensation calculations for halogens

Experimental and/or theoretical studies of halogen condensation kinetics are beyond the scope of this paper. Following Fegley & Lewis (1980) we did condensation calculations assuming reactions of halogens with pre-existing condensates were kinetically inhibited. In this case the rapid accretion of smaller grains into larger bodies prevents the material within the bodies from reacting with nebular gas (cf. Turekian & Clark 1969). The result is a disequilibrium condensation sequence that is path dependent. It is comparable to the cloud condensation sequence in brown dwarfs where observation of monatomic alkali gas lines shows alkali-bearing feldspars did not condense at greater depths and higher temperatures (Lodders 1999b).

Table 18 shows results of our kinetic inhibition (aka chemical disequilibrium) condensation calculations for halogens and alkalis at $10^{-4}$ bar total pressure in the solar nebula. Sodium sulfide $Na_2S$ is the first alkali-bearing condensate at 776 K and removes nearly all Na from the gas before NaF forms at 650 K. Either halite (NaCl, 626 K) or sylvite (KCl, 622 K) is the first chlorine-bearing condensate. Bromine condenses as KBr, either dissolved in solid solution in halite or as pure KBr. Iodine condenses in RbI and CsI. The 50% condensation temperatures for the halogens are 638 K fluorine, 612 K chlorine, 498 K bromine, and 450 K iodine. This sequence is not the extreme endmember studied by Fegley & Lewis (1980) because we allowed NaCl to react with nebular gas at lower temperatures. If we completely prevented any reaction of pre-existing condensates and nebular gas, only NaF forms and Cl, Br, and I remain in the gas until ammonium halides form at much lower temperatures: $NH_4Cl$ 278 K, $NH_4Br$ 273 K, and $NH_4I$ 259 K. The 50% condensation temperatures are then 262 K chlorine, 259 K bromine, and 246 K iodine. Zolotov & Mironenko (2007) also modeled kinetic inhibition of chlorine condensation and computed $HCl\cdot 3H_2O$ formation at 140-160 K.

**Table 18. Halogen condensation Temperatures at $10^{-4}$ bar: Kinetic Inhibition Model**

| Phase | T/K | Notes |
|---|---|---|
| $Na_2S$ start | 776 | S/Na ~ 7.3, consumes all Na, Na ~97.5% in $Na_2S$ at 643 K |
| Na 50% | ~762 | |
| NaF start | 650 | but Na(g) ~ 1,430 atoms > F 1270 atoms |
| 50% F | ~638 | |
| NaCl start | 626 | pure halite, the NaCl used up some $Na_2S$, Br/Cl ~ $2 \times 10^{-3}$; a(KCl) = 0.78 |
| KCl start | 622 | pure sylvite (KCl), 1st K-condensate |
| Cl 50% | ~613 | Chlorine in KCl + NaCl |
| K 50% | ~612 | Potassium in KCl |
| KBr in NaCl | 616 | 50% Br condensed in NaCl ideal solution, but expect activity coefficient > 1, both K and Br larger ions than Na and Cl |
| KBr start | 517 | pure KBr |
| RbBr start | 500 | pure RbBr |
| Br 50% | 498 | Bromine in KBr + RbBr |
| Rb 50% | ~492 | Rubidium in RbCl |
| RbI start | 460 | pure RbI |
| CsI start | 451 | pure CsI |
| RbI | 460 | |
| I 50% | 450 | Iodine in RbI + CsI |
| Cs 50% | 446 | Cesium in CsI |

## 6. Conclusions

We reviewed available analytical procedures and concentrations of halogens in chondrites and point out analytical challenges associated with CI-chondrites. Historically, the halogen abundances have been quite uncertain and unfortunately remain so. We still need reliable measurements from large, representative and well-homogenized CI-chondrite samples. Our recommended values are close to previously accepted values. Average concentrations by mass for CI-chondrites are F = 92±20 ppm, Cl = 717±110 ppm, Br = 3.77±0.90 ppm, and I = 0.77±0.31 ppm. The meteoritic abundances on the atomic scale normalized to N(Si) =106 are N(F) = 1270±270, N(Cl) = 5290±810, N(Br)= 12.3±2.9, and N(I) = 1.59±0.64. The meteoritic logarithmic abundances scaled to present-day photospheric abundances with log N(H) = 12 are A(F)= 4.61±0.09, A(Cl)=5.23±0.06, A(Br) = 2.60±0.09, and A(I) = 1.71±0.15. These are our recommended present-day solar system abundances. These are compared to the present-day solar values derived from sun-pots of N(F) = 776±260, A(F) = 4.40±0.25 , and N(Cl) = 5500±810, A(Cl) = 5.25±0.12. The recommended abundances for F and Cl based on meteorites are consistent with halogen abundances in other stars and other astronomical environments. Comparison data for Br are sparse, and unavailable for I. Additional Br and I analyses in different astronomical environments are needed. Updated equilibrium 50% condensation temperatures at a total pressure of $10^{-4}$ bar for a solar composition gas are 713 K(F), 427 K (Cl), 392 K (Br) and 312 K (I). We also give condensation temperatures considering solid-solutions as well as kinetic inhibition effects. Condensation temperatures computed with lower halogen abundances do not represent the correct condensation temperatures from a solar composition gas.

## Acknowledgements

This paper was delayed due to many circumstances beyond our control. We thank H. Palme, and an anonymous reviewer for constructive reviews. We acknowledge the review by H. Busemann strongly arguing that all meteoritic samples are contaminated with halogens. We thank Mark Kendrick for helpful comments, and the editor in chief, Astrid Holzheid for patience during the revision process. Work supported by NSF-AST 1517541 and the McDonnell Center for the Space Sciences.